\documentclass[epj]{webofc}
\usepackage[utf8]{inputenc}
\usepackage[varg]{txfonts}   
\usepackage{booktabs}
\usepackage{xcolor}
\definecolor{darkred}{rgb}{0.4,0.0,0.0}
\definecolor{darkgreen}{rgb}{0.0,0.4,0.0}
\definecolor{darkblue}{rgb}{0.0,0.0,0.4}
\usepackage[bookmarks,linktocpage,colorlinks,
    linkcolor = darkred,
    urlcolor  = darkblue,
    citecolor = darkgreen]{hyperref}
\usepackage{slashed}
\usepackage{MnSymbol}
    
\newcommand{\ba}{\begin{eqnarray}}
\newcommand{\ea}{\end{eqnarray}}
\newcommand{\be}{\begin{equation}}
\newcommand{\ee}{\end{equation}}
\newcommand{\bd}{\begin{displaymath}}
\newcommand{\ed}{\end{displaymath}}
\newcommand{\bi}{\begin{itemize}}
\newcommand{\ei}{\end{itemize}}

\newcommand{\nn}{\nonumber}
\newcommand{\MS}{\overline{\mathrm{MS}}}

\newcommand{\muRI}{\mu_{\mathrm{RI}}}

\usepackage{subfigure}
\wocname{EPJ Web of Conferences}
\woctitle{Lattice2017}
\begin{document}
%
\selectlanguage{english}
\title{%
Recent progress in applying lattice QCD to kaon physics
}
\author{%
    \firstname{Xu} \lastname{Feng}\inst{1,2,3}\fnsep\thanks{Speaker, \email{pkufengxu@gmail.com}} 
}
\institute{%
School of Physics and State Key Laboratory of Nuclear Physics and Technology, Peking University, Beijing 100871, China
\and
Collaborative Innovation Center of Quantum Matter, Beijing 100871, China
\and
Center for High Energy Physics, Peking University, Beijing 100871, China
}
\abstract{%
 Standard lattice calculations in kaon physics are based on the evaluation of matrix elements of local operators between two single-hadron states or a single-hadron state and the vacuum. Recent progress in lattice QCD has gone beyond these standard observables. I will review the status and prospects of lattice kaon physics with an emphasis on non-leptonic $K\to\pi\pi$ decay and long-distance processes including $K^0$-$\overline{K^0}$ mixing and rare kaon decays.
}
\maketitle
\section{Introduction}\label{intro}

Since the discovery of kaons, the kaon physics plays a key role in the building of the Standard Model.
The main mission for lattice QCD in kaon physics is to evaluate the low-energy
hadronic effects to test the Standard Model parameters or to constrain on new
physics. Lattice QCD has been successful for the calculations of the observables
such as the pion and kaon decay constants $f_{K^\pm/\pi^\pm}$, the
$K\to\pi\ell\nu$ semileptonic form factor $f_+(0)$ and the neutral kaon mixing 
parameter $B_K$. We refer these observables as ``standard''. Their relevant hadronic 
matrix elements have only one local operator insertion. The initial and final states 
involve at most one stable hadron. Besides, the spatial momenta carried by 
initial/final-state particles are much smaller than the ultraviolet lattice cutoff $1/a$, with
$a$ the lattice spacing. 
These standard observables can be computed with high statistical precision and 
controlled systematic errors using lattice QCD simulations.

Many interesting observables in kaon physics, however, are not ``standard''. One
example is the calculation of $K\to\pi\pi$ decay where the final state involves
multiple hadrons. Another example is the evaluation of the long-distance
contributions to flavor changing processes such as the calculation of the real
and imaginary parts of $K^0$-$\overline{K^0}$ mixing amplitudes, which are
related to the $K_L$-$K_S$ mass difference $\Delta M_K$ and the indirect CP
violating parameter $\epsilon$. Rare kaon decays including $K\to\pi\nu\bar{\nu}$
and $K\to\pi\ell^+\ell^-$ also belong to this category. As these transitions
proceed via the second-order weak interaction, the calculations would involve
the construction of 4-point correlation function and the treatment of non-local 
matrix elements with two effective operator insertions. To tackle such quantities, 
one needs to develop new techniques.

In this report, I will first summarize the lattice QCD calculation of standard observables. 
They include $f_{K^\pm/\pi^\pm}$, $f_+(0)$, and inclusive $\tau\to s$ decay. 
All these quantities are related to the determination of the 
Cabibbo–Kobayashi–Maskawa (CKM) matrix element $|V_{us}|$. 
I will also discuss the current status for the computation of $B_K$, based on both Standard 
Model and beyond. In the second part of this report, I will review the 
lattice calculations of non-standard observables such as $K\to\pi\pi$ decay, 
$K^0$-$\overline{K^0}$ mixing and rare kaon decays, presenting both recently-updated 
lattice results and the newly-developed lattice methodology.

\section{Lattice QCD calculation of standard observables}

\subsection{\boldmath$f_+(0)$, $f_{K^\pm}/f_{\pi^\pm}$ and resulting $|V_{us}|$}

According to the average from Flavor Lattice Averaging Group (FLAG), 
updated in Nov. 2016, lattice QCD calculations of $K_{\ell3}$ form factor $f_+(0)$ 
and the ratio of decay constants $f_{K^\pm}/f_{\pi^\pm}$ have reached to 
the precision of 0.28\% and 0.25\%~\cite{Aoki:2016frl}
\be
f_+(0)=0.9706(27),\quad \frac{f_{K^\pm}}{f_{\pi^\pm}}=1.1933(29). 
\ee
Meanwhile, the precision experimental measurements of $K_{\ell3}$ and 
leptonic decays yield the product $|V_{us}|f_+(0)$~\cite{Moulson:2014cra} and the
ratio $|V_{us}/V_{ud}|f_{K^\pm}/f_{\pi^\pm}$~\cite{Moulson:2014cra,Rosner:2015wva}
\be
|V_{us}|f_+(0)=0.2165(4),\quad \left|\frac{V_{us}}{V_{ud}}\right|\frac{f_{K^\pm}}{f_{\pi^\pm}}=0.2760(4).
\ee
Lattice inputs of $f_+(0)$ and $f_{K^\pm}/f_{\pi^\pm}$ together with the experimental 
data give a precise determination of the CKM matrix elements
\be
\label{eq:Vus_lattice}
|V_{us}|=0.2231(7),\quad \left|\frac{V_{us}}{V_{ud}}\right|=0.2313(7).
\ee

In the Standard Model, the CKM matrix is unitary. Most stringent test of 
CKM unitarity is given by the first row condition
\be
|V_u|^2\equiv|V_{ud}|^2+|V_{us}|^2+|V_{ub}|^2=1.
\ee
Using the results of $|V_{us}|$ and $|V_{ud}|$ given in Eq.~(\ref{eq:Vus_lattice}), 
one finds that $|V_u|^2=0.9798(82)$, which has a 2.5~$\sigma$ deviation from CKM unitarity. 
Currently the most precise determination of $|V_{ud}|=0.97420(21)$ is from superallowed 
nuclear $\beta$ decay~\cite{Hardy:2014qxa,Hardy:2016vhg}. Using $|V_{us}|$ from $K_{\ell3}$ decay 
and $|V_{ud}|$ from nuclear $\beta$ decay sharpens the unitarity test with a much 
smaller uncertainty. However, the deviation is still around 2.4~$\sigma$, as shown 
in the second line of Eq.~(\ref{eq:Vus0}). 
If using $|V_{us}/V_{ud}|$ from leptonic decays and $|V_{ud}|$ from nuclear $\beta$ decay, 
then the result confirms the CKM unitarity; see the third line of Eq.~(\ref{eq:Vus0}). 
The above tests of CKM unitarity are put together here for a comparison
\be
\label{eq:Vus0}
|V_u|^2=
\begin{cases}
0.9798(82), & K_{\ell3} + \mbox{leptonic decays},\\
0.9988(5), & K_{\ell3} + \mbox{nuclear $\beta$ decay},\\
0.9998(5), & \mbox{leptonic + nuclear $\beta$ decay}.
\end{cases}
\ee

\begin{figure}[thb]
  \centering
  \includegraphics[width=0.8\textwidth,clip]{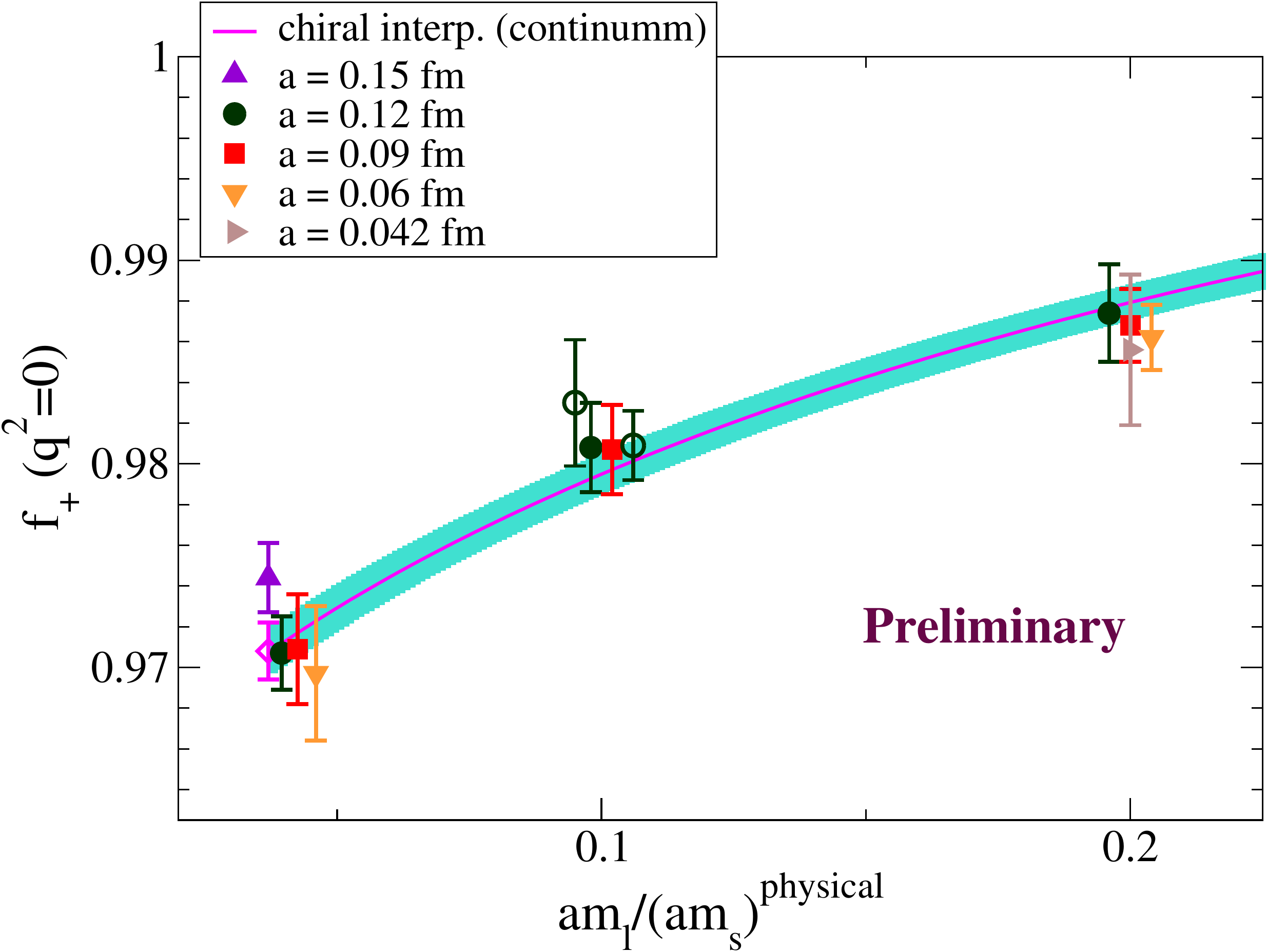}
  \caption{Form factor $f_+(0)$ vs. light-quark mass from Fermilab Lattice-MILC collaboration. 
The calculation is performed at 5 lattice spacings 0.15, 0.12, 0.09, 0.06 and 0.042~fm, 
including 4 ensembles with physical pion mass.
Open green symbols correspond to different volumes for a=0.12~fm and $m_l=0.1\,m_s$. 
The solid magenta line is the (preliminary) interpolation in the light-quark mass, 
keeping the strange-quark mass $m_s$ equal to its physical value, and turning off all discretization effects. 
The magenta diamond is the corresponding interpolation at the physical point. 
Data at the same light-quark mass but different lattice spacing are off-set horizontally.}
  \label{fig:f+_Fermilab_MILC}
\end{figure}

To clarify the 2.x~$\sigma$ deviation in the unitarity test, it is important to 
reduce the uncertainty from the lattice QCD determination of $f_+(0)$. One of 
the recent updates for $f_+(0)$ is from Fermilab Lattice-MILC collaboration. 
HISQ fermions on 2+1+1 flavor MILC configurations are used in the calculation 
and preliminary results are shown in Fig.~\ref{fig:f+_Fermilab_MILC}. Compared 
to their report last year~\cite{Gamiz:2016bpm}, more lattice ensembles are used in the analysis. Employing
4 ensembles at the physical pion mass and 2 ultra-fine lattice spacings allows them to reduce the statistical error to 0.14\%. 
At 0.12 fm and $m_l/m_s=0.1$, they use three different volumes. Three volumes 
together with one-loop chiral perturbation theory (ChPT)~\cite{Bernard:2017scg} allow 
for a good estimate of the finite-volume effects. 
After chiral and continuum extrapolation, the total uncertainty is expected to be 
reduced to 0.2\%, which is close to the current experimental uncertainty~\cite{Moulson:2014cra}. 

Another lattice calculation is recently reported by JLQCD collaboration~\cite{Aoki:2015pba,Aoki:2017spo}. 
In their calculation, the chiral symmetry is exactly preserved by using the overlap quark action, 
which enables a direct comparison of the lattice data with ChPT and hence a determination of 
relevant low energy constants within NNLO ChPT.
A reasonable agreement between lattice results for the slope $df_+(q^2)/dq^2$ (at $q^2 = 0$) and experiment is observed, 
although the error is still large due to the high cost of the usage of overlap fermion.

\subsection{\boldmath$\tau$ inclusive decay and $|V_{us}|$}

\begin{figure}[tp]
   \centering
             {\includegraphics[width=0.52\textwidth,clip]{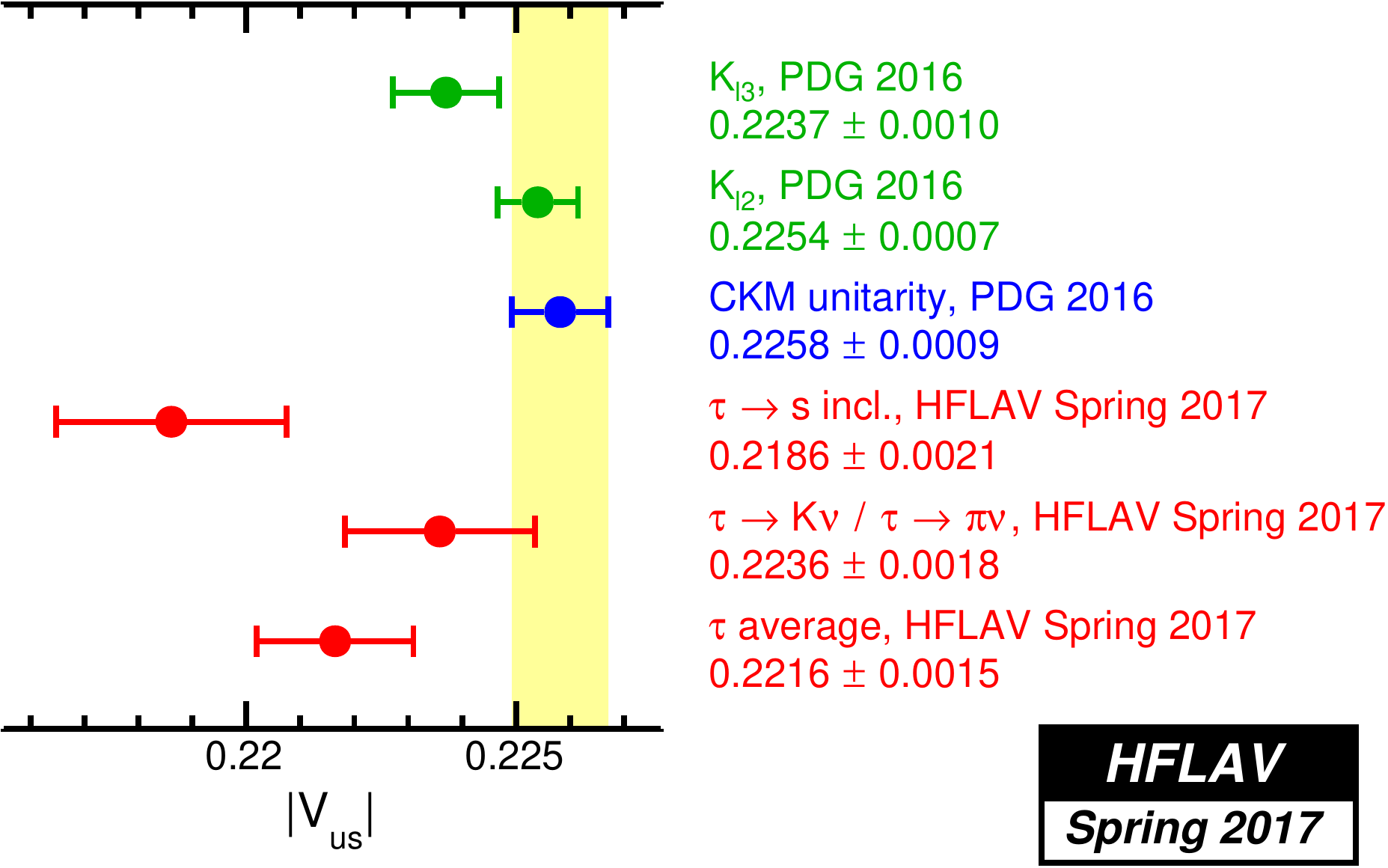}}\hfill
             {\includegraphics[width=0.43\textwidth,clip]{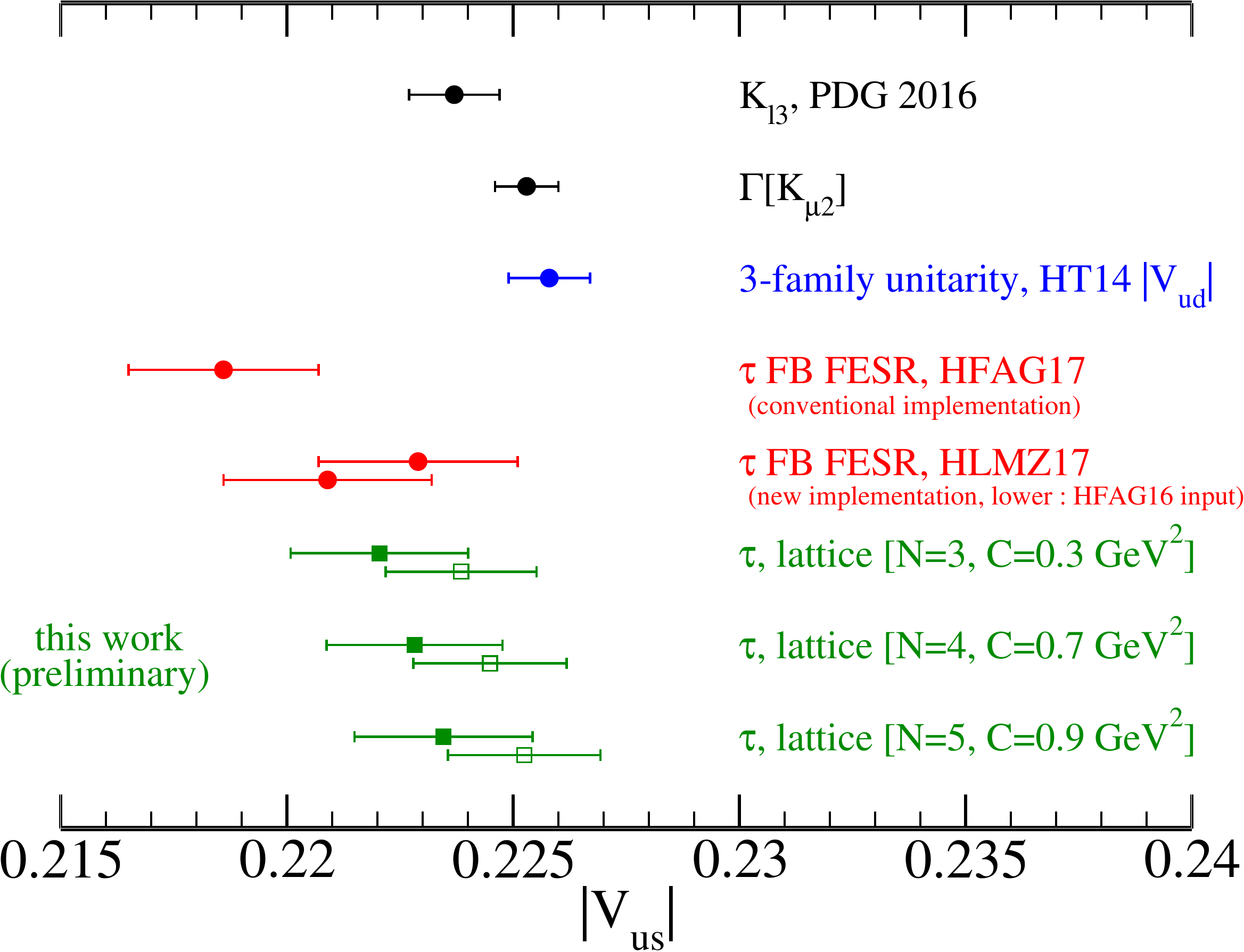}}
 \caption{Left: HFLAV summary of $|V_{us}|$. For the $\tau\to s$ inclusive decay, 
 there is a 3.2~$\sigma$ deviation from CKM unitarity. Right: New implementations 
 for $\tau\to s$ inclusive decay. The read circle data points, denoted as ``$\tau$ FB FESR, HLMZ17'', 
 show the improvement by fitting the OPE effective condensates to the experimental measurements 
 and lattice QCD data~\cite{Hudspith:2017vew}. The green square data points show 
 the improvement using $W(s)$ in Eq.~(\ref{eq:weight_func}) and using lattice HVPs for 
 the residues of the integral~\cite{Ohki}. Both improvements shed the light on the 
 resolution of the puzzle from the $\tau\to s$ inclusive decay.}
   \label{fig:Vus_summary}
\end{figure}

The average of $|V_{us}|$ is summarized and updated on Spring 2017 by Heavy Flavor 
Averaging group (HFLAV)~\cite{Amhis:2016xyh}; see the left panel of Fig.~\ref{fig:Vus_summary}, 
where $f_+(0)$ and $f_{K^\pm}/f_{\pi^\pm}$ take the value from PDG 2016~\cite{Patrignani:2016xqp}. 
The result from leptonic decays shows consistency with CKM unitarity, while the one 
using $K_{\ell3}$ decays has a $\sim$ 2 $\sigma$ deviation from CKM unitarity. 
The largest discrepancy happens for the case using the $\tau\to s$ inclusive decay, 
where a 3.2~$\sigma$ deviation from CKM unitarity is observed.

To explore the discrepancy, the main quantity of interest is the ratio of the decay rates 
\be
R=\frac{\Gamma(\tau\to s\mbox{-hadrons}\,\nu_{\tau})}{\Gamma(\tau\to e\bar{\nu}_e\nu_\tau)},
\ee 
where $\tau\to s\mbox{-hadrons}\,\nu_{\tau}$ indicates that in the decay the final-state hadrons contain net strange-ness. 
According to the optical theorem, the imaginary part of the hadronic vacuum polarization (HVP) functions can be related to
the $R$-value through~\cite{Braaten:1991qm}
\be
\label{eq:optical_theorem}
\frac{dR}{ds}=\frac{12\pi|V_{us}|^2S_{EW}}{m_\tau^2}\left(1-\frac{s}{m_\tau^2}\right)^2\left[\left(1+2\frac{s}{m_\tau^2}\right)\operatorname{Im}\Pi^{(1)}(s)+\operatorname{Im}\Pi^{(0)}(s)\right],
\ee
where $s$ is the invariant mass square of the final-state hadrons. $S_{EW}$ is a known 
short-distance electroweak correction~\cite{Erler:2002mv}.  $\Pi^{(J)}(s)$ are the 
HVP functions with the superscript $(J)$ corresponding to angular momenta $J=0,1$.

 Once $\operatorname{Im}\Pi^{(J)}(s)$ is known, Eq.~(\ref{eq:optical_theorem}) can be used to determine $|V_{us}|$. 
Since $\operatorname{Im}\Pi^{(J)}(s)$ is generically non-perturbative at small $s$, 
the conventional approach to determine $\operatorname{Im}\Pi^{(J)}(s)$ is to use the dispersion relation~\cite{Braaten:1991qm}
\be
\int_{4m_\pi^2}^{s_0}ds\,W(s)\operatorname{Im}\Pi(s)=\frac{i}{2}\oint_{|s|=s_0}ds\,W(s)\Pi(s),
\ee
where $\operatorname{Im}\Pi(s)$ on the left-hand side can be related to $dR/ds$ and $|V_{us}|$, 
while the integral on the right-hand side can be determined using QCD perturbation theory (pQCD) 
and operator product expansion (OPE). The parameter $s_0$ should be sufficiently large for 
a good convergence of pQCD and the validity of the OPE. $W(s)$ is a weight function.  
 If there is no pole inside the contour, then the integral along the branch cut is equal to 
 the integral on the circle and then $|V_{us}|$ can be determined. 
A difficulty here is that the estimate of high-dimensional OPE terms relies on some assumptions 
and thus contains potentially large systematic effects. Using the conventional approach 
described above, it results in the low value of $|V_{us}|$ shown in the left panel of Fig.~\ref{fig:Vus_summary}~\cite{Gamiz:2004ar}.

An improvement is proposed by Ref.~\cite{Hudspith:2017vew} to use different $s_0$ and weight functions 
$W(s)$ and then study the dependence on $s_0$ and $W(s)$. Through fit, not only $|V_{us}|$, 
but also the OPE effective condensates are fit to experimental measurements (and also lattice QCD data).
With this improvement, the 3.2~$\sigma$ deviation is reduced to 1-2~$\sigma$ level depending on 
using BaBar or 2014 HFAG result for $\mathrm{Br}[\tau^-\to K^-\pi^0\nu_\tau]$
  \be
 |V_{us}|=
 \begin{cases}
 0.2229(22) & \mbox{using BaBar $\tau^-\to K^-\pi^0\nu_\tau$, $3.2~\sigma$ $\to$
 $1.2~\sigma$}, \\
 0.2204(23) & \mbox{using HFAG $\tau^-\to K^-\pi^0\nu_\tau$, $3.2~\sigma$ $\to$ $2.2~\sigma$}.
 \end{cases}
\ee
These results are plotted on the right-panel of Fig.~\ref{fig:Vus_summary}, denoted as ``$\tau$ FB FESR, HLMZ17''.

Another new approach proposed by H. Ohki et. al.~\cite{Ohki} is to let $s_0\to\infty$ and 
use the weight function containing the pole structures
\be
\label{eq:weight_func}
W(s)=\prod_{k=1}^{N}\frac{1}{s+Q_k^2},
\ee
where the $N$ different poles $Q_k^2$  are spanned by a spacing $\Delta= 0.2/(N-1)$ GeV$^2$, with  
a center point called $C$. 
Once $W(s)$ is given, 
the contour of the integral is equal to the residues of the poles, which can be determined 
using lattice HVPs. Thus the value of $|V_{us}|$ can be determined accordingly. 
The strategy to choose $Q_k^2$ is that it should not 
be too large to suppress the contribution from pQCD and OPE at $s>m_\tau^2$. It should 
not be too small to avoid large statistical error from lattice HVPs. 
The realistic calculation is performed using $N_f=2+1$ M\"obius domain wall fermions 
at near-physical pion mass with the lattice spacings $a^{-1}=1.73$ and $2.36$ GeV and 
the lattice volume $V=(\mbox{5 fm})^3$. The corresponding results are summarized
in the right panel of Fig.~\ref{fig:Vus_summary}, denoted by the green square data points 
(The filled-square points are generated using $\tau\to K\nu_\tau$ data 
for the $K$ pole, while the open-square ones use the $K_{\mu2}$ data as input). 
Using different $N$ and $C$, the lattice calculation shows the consistent results. 
Besides, all the data points are systematically larger than the conventional value 
of $|V_{us}|$. At $N=4$ and $C=0.7$ GeV$^2$, $|V_{us}|$ is obtained as
\be
|V_{us}|=
\begin{cases}
0.2228(21), & \mbox{using $\tau\to K\nu_\tau$ input for the $K$ pole} \\
0.2245(16), & \mbox{using $K_{\mu2}$ input for the $K$ pole}.
\end{cases}
\ee
  Both improvements proposed by Refs.~\cite{Hudspith:2017vew} and \cite{Ohki}
shed the light on the resolution of the puzzle from the $\tau\to s$ inclusive decay.

\subsection{\boldmath Neutral-kaon mixing parameter $B_K$ based on Standard Model and beyond} 

The parameter $B_K$ is related to the CP violating part of $K^0$-$\overline{K^0}$ mixing and thus 
short-distance dominated. Using OPE, the effective Hamiltonian $\mathcal{H}_{\mathrm{eff}}^{\Delta S=2}$ 
can be written as a product of the Wilson coefficient $C(\mu)$ and the $\Delta S=2$ local operator $Q^{\Delta S=2}(\mu)$
\be
\mathcal{H}_{\mathrm{eff}}^{\Delta S=2}=\frac{G_F^2M_W^2}{16\pi^2}C(\mu)Q^{\Delta S=2}(\mu),
\ee
with $G_F$ the Fermi constant and $M_W$ the $W$-boson mass.
It is a convention to use the parameter $\epsilon$ as a measure of indirect CP violation,
\be
\label{eq:epsilon_def}
\epsilon=\frac{A(K_L\to(\pi\pi)_{I=0})}{A(K_S\to(\pi\pi)_{I=0})},
\ee
with the initial state given by $K_{L/S}$ particle and the final state having total isospin zero.
The contribution from $\mathcal{H}_{\mathrm{eff}}^{\Delta S=2}$ serves as a dominant contribution to $\epsilon$
\be
\label{eq:epsilon}
\epsilon=\exp(i\phi_{\epsilon})\sin(\phi_\epsilon)\left[\frac{\operatorname{Im}[M_{\bar{0}0}^{\mathrm{SD}}]}{\Delta M_K}
+\frac{\operatorname{Im}[M_{\bar{0}0}^{\mathrm{LD}}]}{\Delta M_K}+\frac{\operatorname{Im}[A_0]}{\operatorname{Re}[A_0]}\right],\quad M_{\bar{0}0}^{\mathrm{SD}}=\langle\overline{K^0}|\mathcal{H}_{\mathrm{eff}}^{\Delta S=2}|K^0\rangle.
\ee
Here the angle $\phi_\epsilon\equiv\operatorname{arctan}(-2\Delta M_K/\Delta
\Gamma_K)\approx 43.52(5)^\circ$~\cite{Patrignani:2016xqp}, with $\Delta M_K=M_{K_L}-M_{K_S}$ and $\Delta
\Gamma_K=\Gamma_{K_L}-\Gamma_{K_S}$. $M_{\bar{0}0}^{\mathrm{LD}}$ indicates the
long-distance contribution to $\epsilon$ and $A_0$ is the $K^0\to(\pi\pi)_{I=0}$ amplitude. 
Both $M_{\bar{0}0}^{\mathrm{LD}}$ and $A_0$ only make few-percent contributions to $\epsilon$. 
The progress in lattice QCD calculation of these two quantities will be discussed later. Here we only focus on the 
$M_{\bar{0}0}^{\mathrm{SD}}$.

Within Standard Model, there is only one $\Delta S=2$ operator with $V-A$ structure 
\be
Q^{\Delta
S=2}=\left[\bar{s}_\alpha\gamma_\mu(1-\gamma_5)d_\alpha\right]\left[\bar{s}_\beta\gamma_\mu(1-\gamma_5)d_\beta\right],
\ee
where the subscripts $\alpha$ and $\beta$ denote the color indices.
 For beyond-Standard-Model theories, 4 other operators are possible
\ba
&&Q_2^{\Delta S=2}=\left[\bar{s}_\alpha(1-\gamma_5)d_\alpha\right]\left[\bar{s}_\beta(1-\gamma_5)d_\beta\right],
\quad Q_3^{\Delta S=2}=\left[\bar{s}_\alpha(1-\gamma_5)d_\beta\right]\left[\bar{s}_\beta(1-\gamma_5)d_\alpha\right],
\nn\\
&&Q_4^{\Delta S=2}=\left[\bar{s}_\alpha(1-\gamma_5)d_\alpha\right]\left[\bar{s}_\beta(1+\gamma_5)d_\beta\right],
\quad Q_5^{\Delta S=2}=\left[\bar{s}_\alpha(1-\gamma_5)d_\beta\right]\left[\bar{s}_\beta(1+\gamma_5)d_\alpha\right].
\ea
The neutral-kaon mixing parameter $B_K$ and $B_i$ in the $\MS$ scheme are defined as
\ba
&&B_K(\mu)=\frac{\langle\overline{K^0}|Q^{\Delta S=2}(\mu)|K^0\rangle}{\frac{8}{3}f_K^2M_K^2},
\nn\\
&&B_i(\mu)=\frac{\langle\overline{K^0}|Q_i^{\Delta S=2}(\mu)|K^0\rangle}{N_i\langle\overline{K^0}|\bar{s}\gamma_5d|0\rangle\langle0|\bar{s}\gamma_5d|K^0\rangle},\quad\{N_2,\cdots,N_5\}=\{-5/3,1/3,2,2/3\},
\ea
where $\mu$ is the renormalization scale, $f_K$ the kaon decay constant and $M_K$ the kaon mass. 
Given the anomalous dimension $\gamma(g)$, the renormalization group independent $B$ 
parameter $\hat{B}_K$ is related to $B_K(\mu)$ by the formula
\be
\hat{B}_K=\left(\frac{\bar{g}(\mu)^2}{4\pi}\right)^{-\gamma_0/(2\beta_0)}\exp\left\{\int_0^{\bar{g}(\mu)}dg\left(\frac{\gamma(g)}{\beta(g)}+\frac{\gamma_0}{\beta_0g}\right)\right\}B_K(\mu).
\ee
For the Standard Model $\hat{B}_K$, the lattice calculation has reached a precision of 
1.3\% for 2+1 flavor calculation~\cite{Aoki:2016frl}
\be
\hat{B}_K=0.763(10).
\ee
For beyond-Standard-Model $B_i(\mu)$ at the $\MS$ scale $\mu=3$ GeV, the uncertainties of 2+1 flavor 
lattice results are about 2-5\%~\cite{Aoki:2016frl}
\be
B_2=0.502(14),\quad B_3=0.766(32),\quad B_4=0.926(19),\quad B_5=0.720(38).
\ee

\begin{figure}[tp]
   \centering
             {\includegraphics[width=0.55\textwidth,clip]{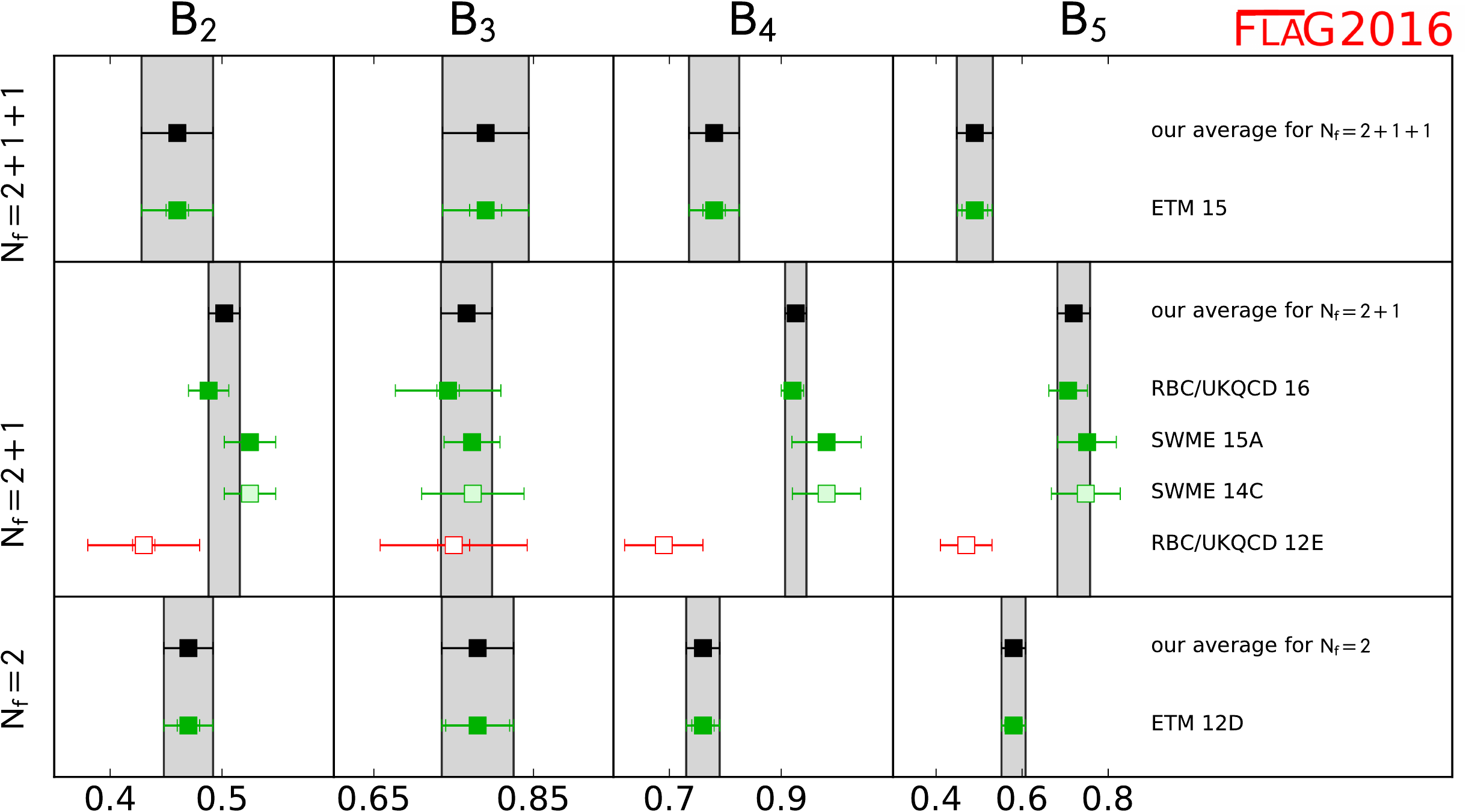}}\hfill
             {\includegraphics[width=0.4\textwidth,clip]{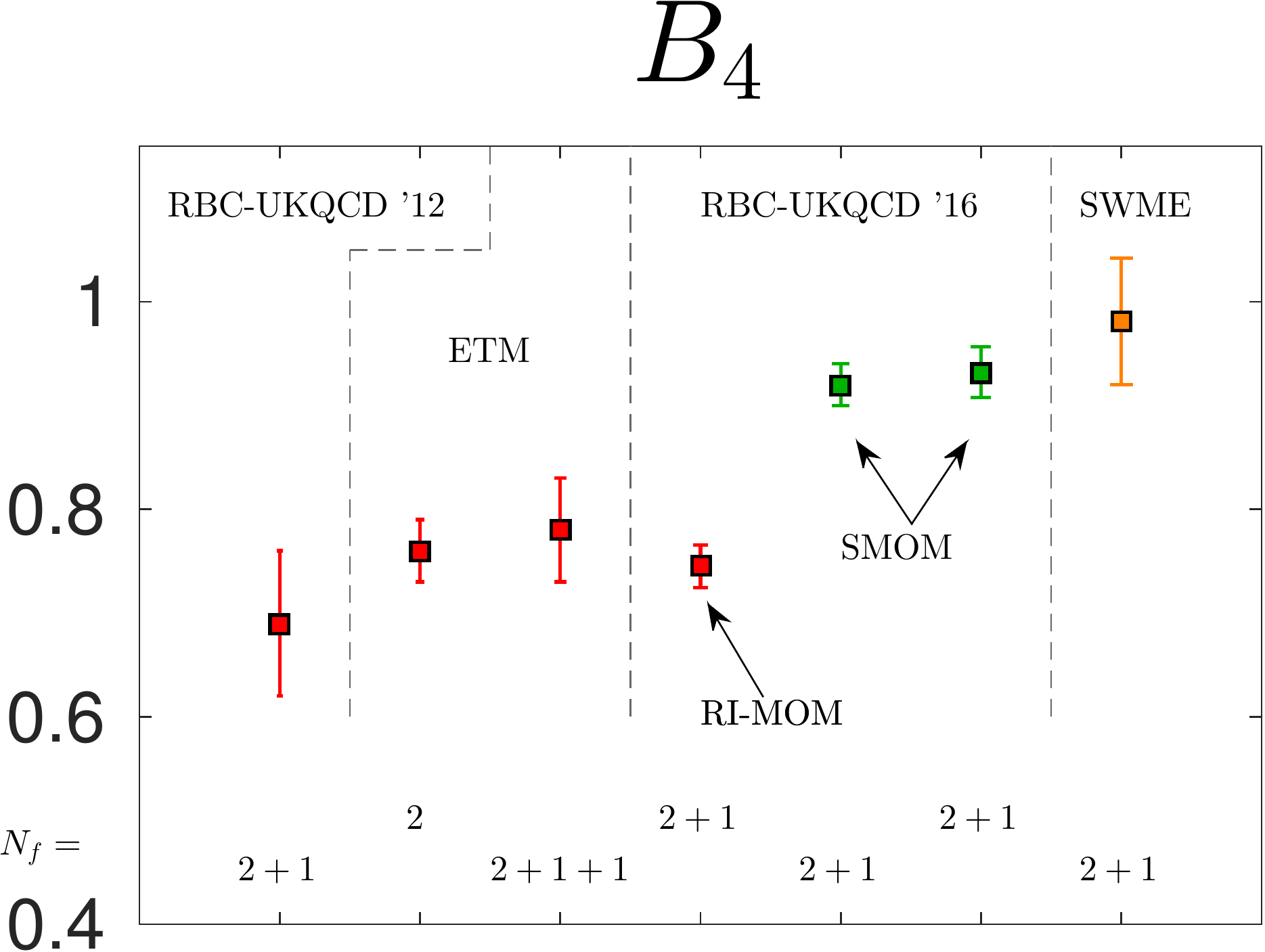}}
 \caption{Left: FLAG summary of $B_i(\mu)$ at $\mu=3$ GeV~\cite{Aoki:2016frl}. 
     There are clear discrepancies in $B_4$ and $B_5$ from different groups. 
     Right: Comparison for $B_4$ using RI-MOM and RI-SMOM non-perturbative 
	renormalization and 1-loop perturbative renormalization. 
     The four red data points use RI-MOM scheme while the two green ones use RI-SMOM scheme and the orange 
     one uses 1-loop lattice perturbation theory.
 All RI-MOM results are compatible among different groups. However, they
 are systematically smaller than the RI-SMOM and 1-loop perturbative results.}
   \label{fig:B_BSM}
\end{figure}
 
The results for $B_i(\mu)$ from various groups are summarized by FLAG~\cite{Aoki:2016frl} 
on the left panel of Fig.~\ref{fig:B_BSM}. There are clear discrepancies in $B_4$ and $B_5$ from different groups.
To resolve these discrepancies, RBC-UKQCD collaboration undertakes a study using both RI-MOM and 
RI-SMOM renormalization~\cite{Garron:2016mva,Boyle:2017skn,Garron}. The calculation is performed using 
$N_f=2+1$ flavor domain wall fermion at $M_\pi\approx300$ MeV and two lattice spacings $a=0.08$ and 0.11 fm. 
The corresponding results are shown by the three data points below the legend of ``RBC-UKQCD '16'' on the right panel of Fig.~\ref{fig:B_BSM},
where the four 
red data points use RI-MOM scheme while the two green ones use RI-SMOM scheme and the orange 
one uses 1-loop lattice perturbation theory.
Including the new RBC-UKQCD updates, all
RI-MOM results are compatible among different groups. However, these results are systematically smaller than that from the RI-SMOM renormalization and the 1-loop lattice perturbation theory.
Particularly for RI-SMOM calculation, both $({\slashed q},{\slashed q})$ and $(\gamma,\gamma)$-schemes are used and
consistent results (shown by the two green data points) are obtained after conversion to $\MS$ scheme. For $B_5$, the situation is very similar.

According to the study by Ref.~\cite{Aoki:2007xm}, RI-SMOM renormalization
is expected to have smaller infrared contamination
than RI-MOM due to the usage of the non-exceptional momenta. 
The studies by~\cite{Garron:2016mva,Boyle:2017skn,Garron} confirm such expectation and suggest that for $B_4$ and $B_5$ RI-SMOM renormalization should be used 
to fully control the infrared contamination. In Ref.~\cite{Boyle:2017ssm}
an update is reported on the RBC-UKQCD measurement of $B_i(\mu)$, simulated
using $N_f=2+1$ domain wall fermions at the physical quark masses.

\section{Go beyond standard observables}

\subsection{\boldmath $K \to \pi\pi$ decay and direct CP violation}

CP violation is first observed in neutral kaon decays. Under CP transform, 
the $K^0$ state is related to $\overline{K^0}$ state through
\be
CP|K^0\rangle=-|\overline{K^0}\rangle.
\ee
The CP eigenstate can be defined as the combination of $K^0$ and $\overline{K^0}$
\be
K_{\pm}^0=\frac{1}{\sqrt{2}}\left(K^0\mp \overline{K^0}\right),
\ee
with $K_{+/-}^0$ the CP-even/odd state.
 The physical states observed in the experiment are the weak eigenstates $K_S$ and $K_L$. 
 $K_S$ decays into two pions and $K_L$ decays into three pions. By neglecting CP violation, 
 $K_S$ is equal to CP-even state and $K_L$ equal to CP-odd state. In 1964, BNL discovered 
 that $K_L$ is able to decay into two pions, indicating the violation of CP symmetry. 
 This discovery leads to the Nobel prize in 1980.

Since $K_L$ and $K_S$ are not CP eigenstates, one can write them as a mixture of the CP eigenstates
\be
|K_{L/S}\rangle=\frac{1}{\sqrt{1+\bar{\epsilon}}^2}\left(|K_\mp^0\rangle+\bar{\epsilon}|K_\pm^0\rangle\right).
\ee
The parameter $\bar{\epsilon}$ is a measure of the strength of mixing.
 For $K_L\to\pi\pi$ decay, there are two contributions to the CP violation. 
The first part appears as the CP-even component of $K_L$ decays into two pions.
This is called indirect CP violation and described by a parameter $\epsilon$ or in many cases written as $\epsilon_K$.
 $\epsilon$ receives its dominant contribution from $\bar{\epsilon}$ and a small contribution from $A_0$ due to its definition given in Eq.~(\ref{eq:epsilon_def})
\be
\label{eq:epsilon_contribution}
\epsilon=\bar{\epsilon}+i\frac{\operatorname{Im}[A_0]}{\operatorname{Re}[A_0]}.
\ee
 The second contribution is from the CP-odd component of $K_L$, which decays into two pions directly. 
 This is called direct CP violation and denoted as $\epsilon'$.

The experiments measure the decay amplitudes of $K_L\to\pi\pi$ and $K_S\to\pi\pi$ and use the ratio
\be
\eta_{+-}\equiv\frac{A(K_L\to\pi^+\pi^-)}{A(K_S\to\pi^+\pi^-)}\equiv\epsilon+\epsilon',\quad
\eta_{00}\equiv\frac{A(K_L\to\pi^0\pi^0)}{A(K_S\to\pi^0\pi^0)}\equiv\epsilon-2\epsilon'
\ee
to determine the parameter $\epsilon$ and $\epsilon'$.
 Using the experimental measurements of $|\eta_{+-}|$ and $|\eta_{00}|$ as input, PDG quotes~\cite{Patrignani:2016xqp}
\be
\label{eq:epsilon_and_epsilon_prime}
|\epsilon|\approx\frac{1}{3}\left(2|\eta_{+-}+|\eta_{00}|\right)=2.228(11)\times10^{-3},\quad
\operatorname{Re}[\epsilon'/\epsilon]\approx\frac{1}{3}\left(1-\frac{|\eta_{00}|}{|\eta_{+-}|}\right)=1.66(23)\times10^{-3}.
\ee
 $\epsilon$ is at the order of $10^{-3}$ and $\epsilon’$ is even 1000 times smaller. 
Due to its small size, direct CP violation $\epsilon’$ is very sensitive to new physics.

For theoretical simplicity, it is convenient to study the decay amplitudes in the specific isospin channels, $A_0$ and $A_2$,
\be
A(K^0\to(\pi\pi)_I)=A_Ie^{i\delta_I},\quad I=0,2,
\ee
where $\delta_I$ is the strong phase from $\pi\pi$ scattering.
If CP symmetry were protected, then both the amplitudes $A_2$ and $A_0$ are real. To obtain the CP violation, one shall determine
both real and imaginary part of $A_2$ and $A_0$.
The indirect CP violation $\epsilon$ only has a small dependence on $A_0$ as shown by Eq.~(\ref{eq:epsilon_contribution}). 
While for $\epsilon'$, it is sensitive on both real and imaginary part of $A_2$ and $A_0$
\be
\epsilon'=\frac{ie^{i(\delta_2-\delta_0)}}{\sqrt{2}}\frac{\operatorname{Re}[A_2]}{\operatorname{Re}[A_0]}
\left(\frac{\operatorname{Im}[A_2]}{\operatorname{Re}[A_2]}-\frac{\operatorname{Im}[A_0]}{\operatorname{Re}[A_0]}\right).
\ee
The target for lattice QCD calculation is to determine $A_2$ and $A_0$ from first principles.

\begin{figure}
       \centering
       \shortstack{\includegraphics[width=.36\textwidth]{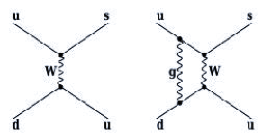}\\
           Current-current \\ $Q_1$, $Q_2$} 
      \shortstack{\includegraphics[width=.19\textwidth]{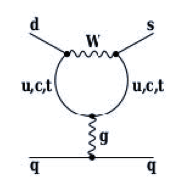}\\
          QCD penguin\\ $Q_3$-$Q_6$} 
     \shortstack{\includegraphics[width=.36\textwidth]{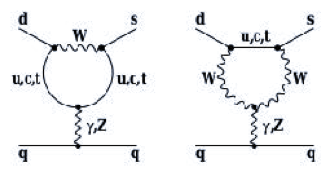}\\
        Electro-weak penguin\\ $Q_7$-$Q_{10}$} 
    \caption{Examples of current-current, QCD penguin and electroweak penguin diagrams in the full theory for $K\to\pi\pi$ decays.}
   \label{fig:operator_classification}
\end{figure}

The weak Hamiltonian for $K\to\pi\pi$ decay is given by a series of $\Delta S=1$ local four-quark operators~\cite{Buchalla:1995vs}
\be
{\mathcal H}_{\mathrm{eff}}^{\Delta S=1}=\frac{G_F}{\sqrt{2}}V_{ud}V_{us}^*\sum_{i=1}^{10}\left[z_i(\mu)+\tau y_i(\mu)\right]Q_i.
\ee
 Here $\tau=-\frac{V_{td}V_{ts}^*}{V_{ud}V_{us}^*}=1.543+0.635i$ is the ratio of CKM matrix elements. 
 $z_i(\mu)$ and $y_i(\mu)$ are known perturbative Wilson coefficients that summarize the short-distance effects. 
 The 10 local four-quark operators $Q_i$ can be matched to three types of diagrams in the full theory, 
 shown in Fig.~\ref{fig:operator_classification} with the notations ``Current-current'', ``QCD penguin'' 
 and ``Electro-weak penguin'', respectively.
$Q_1$, $Q_2$ are current-current operators, which dominate the contribution to $\operatorname{Re}[A_2]$ 
and $\operatorname{Re}[A_0]$. $Q_3$-$Q_6$ are QCD penguin operators, where $Q_6$ dominates 
the contribution to $\operatorname{Im}[A_0]$. $Q_7$-$Q_{10}$ are electroweak penguin operators, 
which dominate the contribution to $\operatorname{Im}[A_2]$.

The most recent updated results for the amplitude $A_2$ are given by RBC-UKQCD 
collaboration~\cite{Blum:2015ywa},  where two ensembles are used, both at physical pion mass 
but with different lattice spacings. The parameters are given in Table~\ref{tab:A2_parameter}.
 After continuum extrapolation the results for $A_2$ are given by
\be
\operatorname{Re}[A_2]=1.50(4)_{\mathrm{stat}}(14)_{\mathrm{syst}}\times10^{-8}\mbox{ GeV},\quad
\operatorname{Im}[A_2]=-6.99(20)_{\mathrm{stat}}(84)_{\mathrm{syst}}\times10^{-13}\mbox{ GeV},
\ee
where $\operatorname{Re}[A_2]$ is consistent with the experimental measurement of 
$\operatorname{Re}[A_2]\approx|A_2|=1.479(3)\times10^{-8}$ GeV obtained from $K^+$ decays. 
For $\operatorname{Im}[A_2]$, it is unknown from experiments. Only lattice QCD provides 
the result. For the $I=2$ $\pi\pi$ scattering phase at $E_{\pi\pi}=M_K$, the lattice calculation yields 
\be
\delta_2=-11.6(2.5)(1.2)^\circ
\ee
which is calculated using L\"uscher's formula~\cite{Luscher:1990ux} and consistent with the phenomenological curve from Ref.~\cite{Schenk:1991xe}.

\begin{table}[thb]
  \small
  \centering
  \caption{Ensembles used in the recent lattice calculation of $A_2$ by RBC-UKQCD collaboration~\cite{Blum:2015ywa}.}
  \label{tab:A2_parameter}
  \begin{tabular}{ccccc}\toprule
  $M_\pi$ (MeV) & $(L/a)^3\times (T/a)$  & $a$ (fm) & $L$  (fm) & $N_{\mathrm{conf}}$ \\\midrule
  139.1(2) & $48^3\times96$ & 0.11 & 5.4 & 76 \\
   139.2(3) & $64^3\times128$ & 0.084 & 5.4 & 40 \\\bottomrule
  \end{tabular}
\end{table}

\begin{figure}
       \centering
       \shortstack{\includegraphics[width=.35\textwidth]{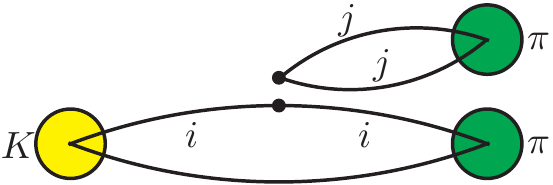}\\
      $C_1$: color diagonal\\
\vspace{0.2cm}
\includegraphics[width=.35\textwidth]{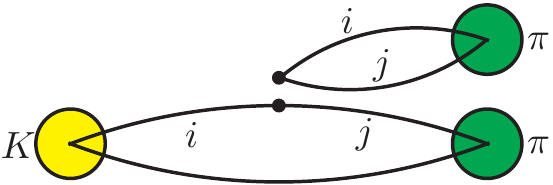} \\
      $C_2$: color mixed } 
\hspace{0.4cm}
     \shortstack{\includegraphics[width=.5\textwidth]{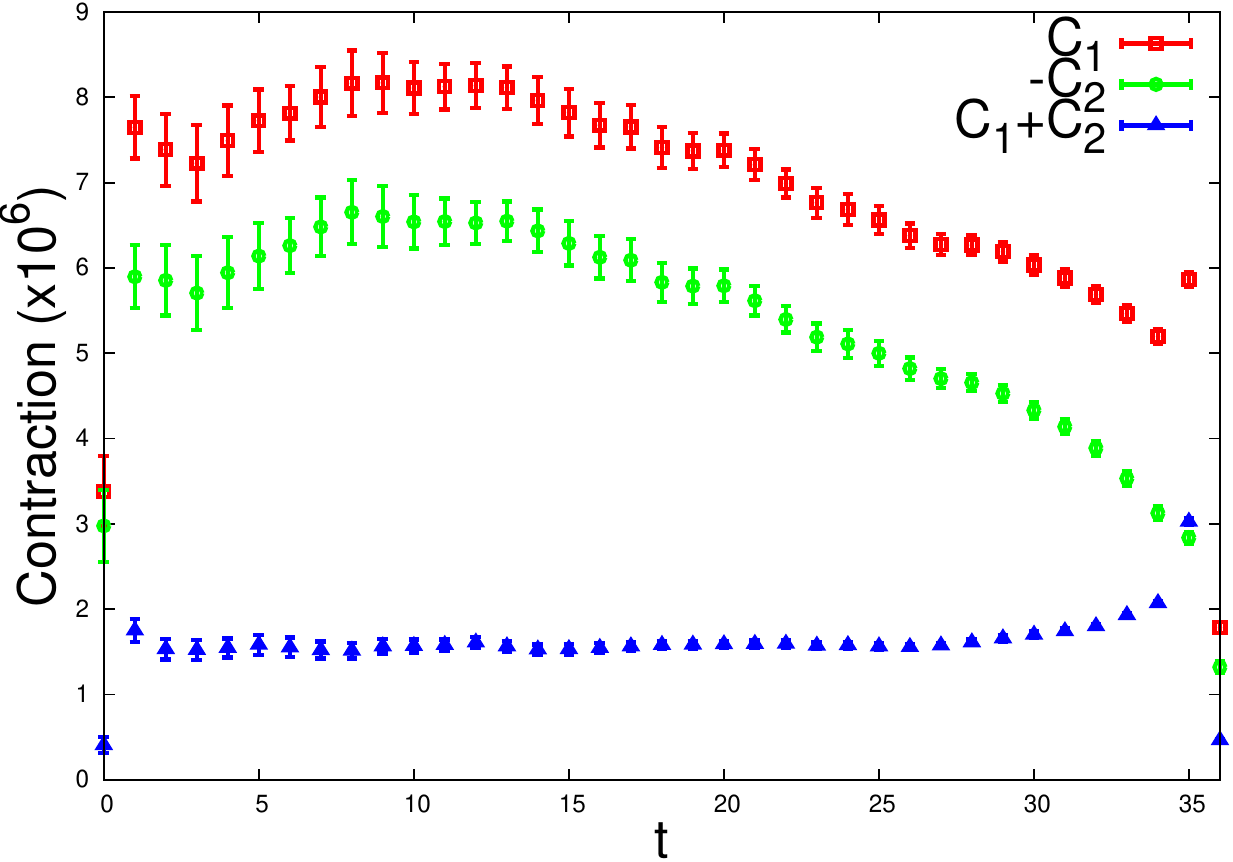}} 
    \caption{Left: Dominant contractions contributing to $\operatorname{Re}[A_2]$. 
    Right: Cancellation of $C_1$ and $C_2$-contributions to $\operatorname{Re}[A_2]$ at the physical pion mass and $a=0.084$ fm.}
   \label{fig:contribution_A2}
\end{figure}

In addition to the determination of $A_2$ and $\delta_2$, another outcome from Ref.~\cite{Blum:2015ywa} is 
the resolution of the puzzle of the $\Delta I=1/2$ rule. According to experimental measurement, the size of 
$A_0$ is about 22.5 times larger than that of $A_2$. It is a more-than-half-century’s puzzle since 
1955~\cite{GellMann:1955jx} on why the amplitudes in the different isospin channels are so much different. 
The Wilson coefficients only account for a factor of 2. The lattice calculation shows that $\operatorname{Re}[A_2]$ 
is dominated by diagrams $C_1$ and $C_2$ in the left panel of Fig.~\ref{fig:contribution_A2}, 
where $C_1$ is color diagonal and $C_2$ color mixed. 
$C_2$ is $1/N$ suppressed relative to $C_1$ with $C_2$ equal to $1/3$ of $C_1$
in leading order QCD perturbation theory.
However, the lattice results in the right panel of Fig.~\ref{fig:contribution_A2} 
shows that $C_2$ is about $-0.7\times C_1$, indicating very strong non-perturbative effects. 
As $\operatorname{Re}[A_2]$ is proportional to $C_1+C_2$, the observation that $C_1$ and $C_2$ 
have opposite signs leads to a significant cancellation between the two terms. 
While for $\operatorname{Re}[A_0]$, the opposite signs lead to an enhancement as 
$\operatorname{Re}[A_0]$ receives an important contribution from $2C_1-C_2$. When 
considering the complete contribution to $\operatorname{Re}[A_0]$, including the 
disconnected diagrams, the size of $\operatorname{Re}[A_0]$ is more enhanced. In total, 
the hadronic matrix elements including the contributions from $C_1$, $C_2$ and 
other diagrams would contribute another factor of $\sim10$. The cancellation between $C_1$ and $C_2$ 
is first observed in an earlier study~\cite{Boyle:2012ys} and is further confirmed by the latest 
calculation of $A_2$~\cite{Blum:2015ywa}. So now the puzzle of $\Delta I=1/2$ rule is resolved 
from first principals. We have also seen a recent study of the $\Delta I=1/2$ rule with 
the scaling of the number of color~\cite{Donini:2016lwz}. 

The more demanding calculation is the $K \to\pi\pi$ decay in the isopsin $I=0$ channel. 
The latest calculation is performed at the physical kinematics $M_\pi=143.1(2.0)$ MeV and $M_K=490(2.2)$ MeV, 
using a $32^3\times64$ lattice volume and a lattice spacing $a=0.14$ fm~\cite{Bai:2015nea}. 
G-parity boundary condition is used and the lattice volume is chosen such that the kaon’s mass 
is equal to the pion-pion’s energy in the ground state. Based on 216 configurations,
the lattice results for $\operatorname{Re}[A_0]$ and $\operatorname{Im}[A_0]$ are reported as
\be
\operatorname{Re}[A_0]=4.66(1.00)_{\mathrm{stat}}(1.26)_{\mathrm{syst}}\times10^{-7}\mbox{ GeV},
\quad
\operatorname{Im}[A_0]=-1.90(1.23)_{\mathrm{stat}}(1.08)_{\mathrm{syst}}\times10^{-11}\mbox{ GeV}.
\ee
Here the real part is consistent with the experimental result: $\operatorname{Re}[A_0]=3.3201(18)\times10^{-7}$ GeV.
The experimental value of $\operatorname{Im}[A_0]$ does not exist, and the knowledge is only from lattice. 
Using L\"uscher's quantization condition~\cite{Luscher:1990ux}, the $I=0$ $\pi\pi$ scattering phase shift is found to be $\delta_0=23.8(4.9)(1.2)^\circ$, 
which is smaller than the value $\delta_0=38.0(1.3)^\circ$, obtained by combining experimental data with 
the Roy equations~\cite{Colangelo:2001df,Colangelo:2015kha}. 
It remains a puzzle for the discrepancy and needs to be understood in the future study. 

Using the lattice results for both $A_0$ and $A_2$, the direct CP violation $\epsilon’/\epsilon$ can be determined:
\be
\label{eq:direct_CP}
\operatorname{Re}[\epsilon'/\epsilon]=0.14(52)_{\mathrm{stat}}(46)_{\mathrm{syst}}\times10^{-3}. 
\ee
There is a 2.1 $\sigma$ deviation from experimental value $\operatorname{Re}[\epsilon'/\epsilon]=1.66(23)\times10^{-3}$~\cite{PDG2014}. 
As the uncertainties of the lattice results are larger than experimental measurement, to confirm whether 
new physics information can be found in the deviation, more accurate lattice calculations are required.

\begin{figure}[thb]
  \centering
  \includegraphics[width=0.8\textwidth,clip]{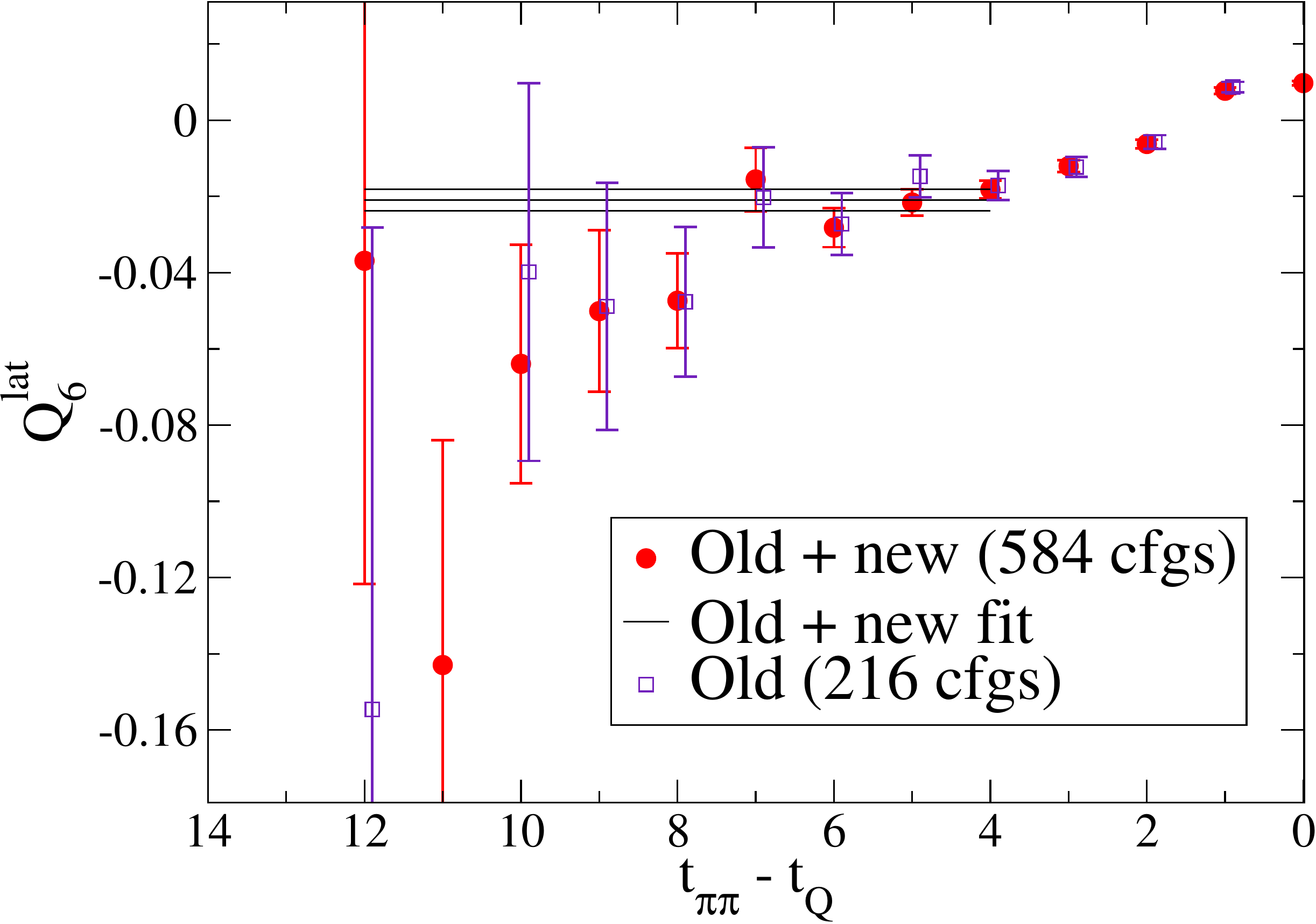}
  \caption{The three-point function for operator $Q_6$ as a function of the time separation between
$Q_6$ and the $\pi\pi$-field insertion $J_{\pi\pi}$, where the time dependence on the $\pi\pi$ energy and 
the kaon mass has been removed. The square data points show the old lattice results based on 216 configurations from Ref.~\cite{Bai:2015nea}.
The circle data points present the updated results with a total of 584 configurations. 
The horizontal lines show the central value and errors from the fit to the updated lattice results.}
  \label{fig:Q6}
\end{figure}

It is reported by C. Kelly~\cite{Kelly} that the statistics of the previous RBC-UKQCD 
calculation has been increased to 584 configurations.
In the lattice calculation, the largest contribution to $\operatorname{Im}[A_0]$ comes from $Q_6$ operator. Fig.~\ref{fig:Q6} shows 
the fit to obtain the matrix element $\langle\pi\pi|Q_6|K\rangle$. When the
statistics increases from $216$ to $584$ configurations, uncertainty decreases 
as expected while the central values remain consistent.
The aim of the RBC-UKQCD $K\to\pi\pi$ program is to reduce the dominant statistical 
error for $\operatorname{Re}[\epsilon'/\epsilon]$ in Eq.~(\ref{eq:direct_CP}) by a factor of 2 within the next year.

Besides for the effort to increase the statistics, there are also efforts for improvements of various systematic effects. 
For example, the $\sigma$ field starts to be added into the calculation to account for the $\sigma\to\pi\pi$ effects 
in the $I=0$ $\pi\pi$ scattering channel. In Ref.~\cite{Christ} N. H. Christ reports on including electromagnetism 
in $K\to\pi\pi$ decay. The $\Delta I=1/2$ rule may make the effects of electromagnetism on $A_2$ $\sim$20 times 
larger than a naive $O(\alpha_e)$ estimate due to the mixing with $A_0$. Such effects will become important 
if the target of the future calculations is to determine $\epsilon'/\epsilon$ with a precision of $\sim10\%$.
In Ref.~\cite{Bruno} M. Bruno presents a non-perturbative calculation of Wilson coefficients even 
including the $W$-boson by using the technique of step scaling. Although the current availability of 
lattice spacings restricts the calculation to unphysically light $W$-bosons with $M_W\sim2$ GeV, 
the calculation opens a new direction in the future to non-perturbatively determine the Wilson coefficients with controlled uncertainties.

In addition to the efforts from RBC-UKQCD collaboration to compute the $K\to\pi\pi$ decay, N. Ishizuka et. al. 
are running a parallel program using the improved Wilson fermion action~\cite{Ishizuka:2015oja}. 
As a first step to verify the possibility of calculations with the Wilson fermion action, 
they consider the decay amplitudes at an unphysical quark mass $M_K\sim 2M_\pi$. 
A large enhancement of the ratio $A_0/A_2$ is found at unphysical quark masses.

\subsection{\boldmath Long-distance contributions to flavor changing process: $\Delta M_K$ and $\epsilon$}

Both the $K_L$-$K_S$ mass difference $\Delta M_K$ and indirect CP violating parameter $\epsilon$ 
are related to the mixing of the  $K^0$ and $\overline{K^0}$. Such mixing is caused by 
the weak interaction as the strangeness differs by 2 in $K^0$ and $\overline{K^0}$. 
The time evolution of the $K^0$-$\overline{K^0}$ mixing system can be given by the equation
\be
i\frac{d}{dt}\begin{pmatrix}K^0(t)\\ \overline{K^0}(t)\end{pmatrix}=
\left[\begin{pmatrix} M_{00} & M_{0\bar{0}} \\ M_{\bar{0}0} & M_{\bar{0}\bar{0}} \end{pmatrix}
-\frac{i}{2}
\begin{pmatrix}
\Gamma_{00} & \Gamma_{0\bar{0}} \\
\Gamma_{\bar{0}0} & \Gamma_{\bar{0}\bar{0}}
\end{pmatrix}
\right]
\begin{pmatrix}
K^0(t)\\ \overline{K^0}(t)
\end{pmatrix}
, 
\ee
where $M$ is the mass matrix and $\Gamma$ the decay width matrix. These $2\times2$ matrices 
are calculated to the $2^{\mathrm{nd}}$ order of the weak interaction and given by
\be
M_{ij}=M_K\delta_{ij}+\langle i|H_W|j\rangle+\mathcal{P}\sumint_\alpha\frac{\langle i|H_W|\alpha\rangle\langle\alpha|H_W|j\rangle}{M_K-E_\alpha},
\quad
\Gamma_{ij}=2\pi\sumint_\alpha\langle i|H_W|\alpha\rangle\langle\alpha|H_W| j\rangle\delta(E_\alpha-M_K). 
\ee
where the indices $i$ and $j$ take the values $0$ and $\bar{0}$. $H_W$ is the $\Delta S=1$ 
weak effective Hamiltonian and $\mathcal{P}$ indicates that the principal part 
should be taken when an integral with a vanishing energy denominator is encountered.

The mass matrix can be diagonalized. By neglecting the effects of CP violation, 
the mass difference $\Delta M_K$ can be given by the real part of $M_{\bar{0}0}$ through
\be
\Delta M_K\equiv M_{K_L}-M_{K_S}=2\operatorname{Re}[M_{\bar{0}0}].
\ee
The parameter $\epsilon$ is related to the imaginary part of $M_{\bar{0}0}$ and 
given explicitly in terms of the short-distance and long-distance part of 
$\operatorname{Im}[M_{\bar{0}0}]$ in Eq.~(\ref{eq:epsilon}).

\begin{figure}
         \centering
\subfigure[$\Delta M_K$: long-distance dominated]{
         \shortstack{\includegraphics[width=.33\textwidth]{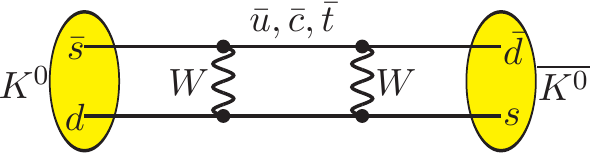}}
	\shortstack{\vspace{0.4cm}\includegraphics[width=.06\textwidth]{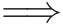}}
	\shortstack{\includegraphics[width=.33\textwidth]{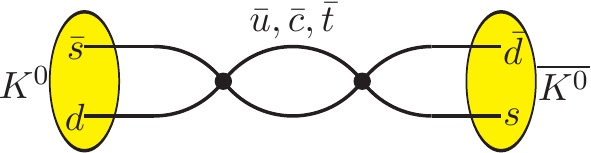}}}
\subfigure[$\epsilon$: short-distance dominated]{
   \shortstack{\includegraphics[width=.33\textwidth]{KL_KS.pdf}}
	\shortstack{\vspace{0.4cm}\includegraphics[width=.06\textwidth]{arrow}}
	\shortstack{\includegraphics[width=.33\textwidth]{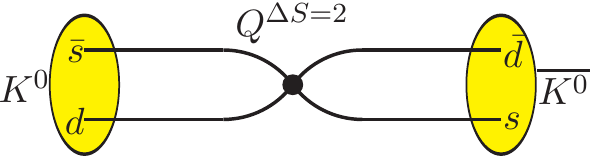}}}
\caption{$K^0$-$\overline{K^0}$ mixing in the full theory. $\Delta M_K$ is related with the 
    CP conserving part of $K^0$-$\overline{K^0}$ mixing and thus long-distance dominated. 
    The process is described by two $\Delta S=1$ operators. $\epsilon$ is related to the CP violating part of 
$K^0$-$\overline{K^0}$ mixing and thus short-distance dominated. The dominant contribution 
is described by a single $\Delta S=2$ operator and the relevant hadronic matrix element 
can be converted to $B_K$. The remaining long-distance contribution below the scale 
of the charm quark mass has been calculated by Ref.~\cite{Bai:2016gzv}.}
\label{eq:K_Kbar_mixing}
     \end{figure}

Both $\Delta M_K$ and $\epsilon$ arise from an amplitude in which two $W$ bosons and 
internal up-type quarks form a loop, shown by Fig.~\ref{eq:K_Kbar_mixing}.  The loop 
integral is proportional to the internal quark mass square $m_q^2$ for $q=u,c,t$. 
As $\Delta M_K$ is related to $\operatorname{Re}[M_{\bar{0}0}]$, it is associated with 
the CP conserving part of $K^0$-$\overline{K^0}$ mixing amplitude. Although the top quark 
loop is enhanced by $m_t^2$, there is a significant suppression from the CKM factor 
$\lambda_t$, where $\lambda_q=V_{qd}V_{qs}^*$. Due to the fact that 
$\operatorname{Re}[\lambda_c^2]\frac{m_c^2}{M_W^2}\gg \operatorname{Re}[\lambda_t^2]\frac{m_t^2}{M_W^2}$, 
the contributions to $\Delta M_K$ are dominated by charm-charm quark loop. As it is sensitive 
to the charm quark mass, the $K_L$-$K_S$ mass difference historically led to the 
predication of the charm quark fifty years ago~\cite{Mohapatra:1968zz,Glashow:1970gm,Gaillard:1974hs}.
For $\epsilon$, it is related to the CP violating part of $K^0$-$\overline{K^0}$ mixing. The charm quark 
contribution is significantly suppressed as $\operatorname{Im}[\lambda_c^2]\ll \operatorname{Re}[\lambda_c^2]$. 
In $\epsilon$, the top-top, top-charm and charm-charm loops compete in size. 
As it contains important top-top loop contribution, $\epsilon$ is sensitive to the Standard Model parameter, $\lambda_t$ or $V_{cb}$.

\begin{figure}
    \centering
     \shortstack{\includegraphics[width=.5\textwidth]{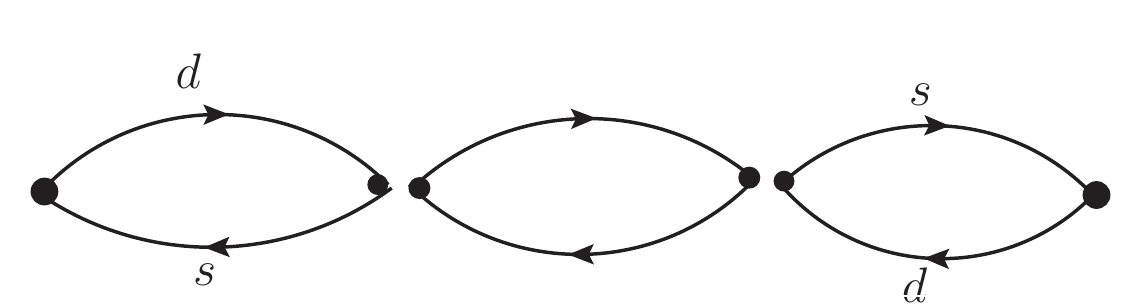}\\Type 1}
     \shortstack{\includegraphics[width=.35\textwidth]{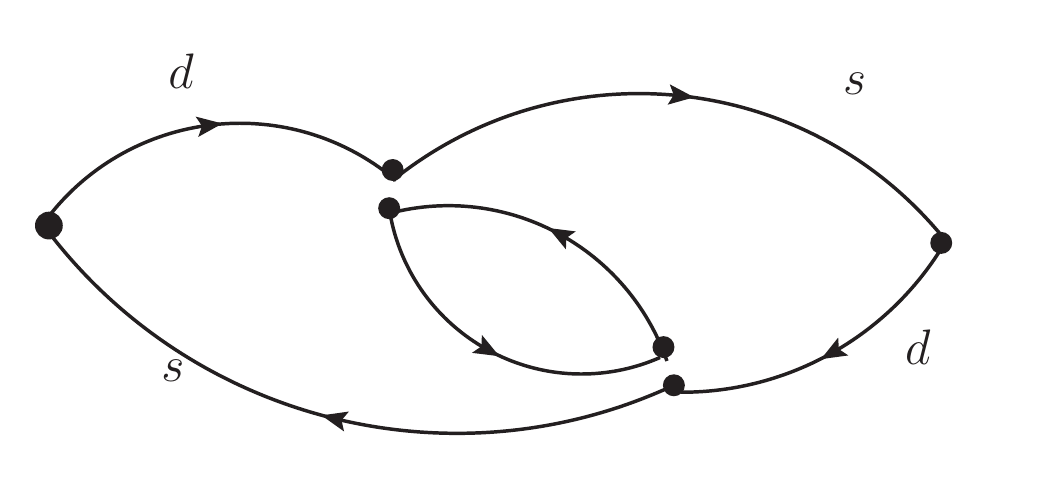}\\Type 2}
     \shortstack{\includegraphics[width=.35\textwidth]{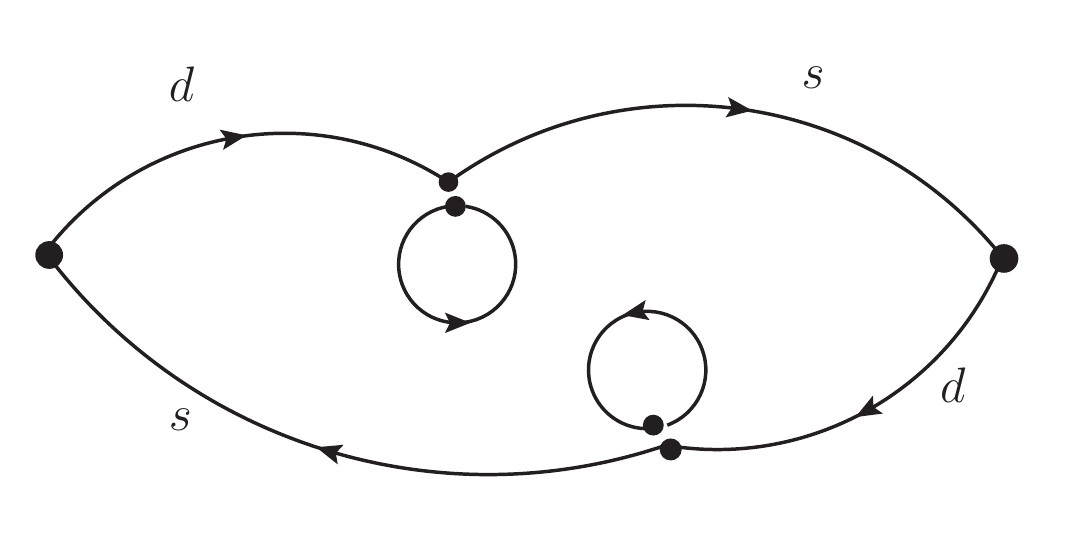}\\Type 3}
     \shortstack{\includegraphics[width=.5\textwidth]{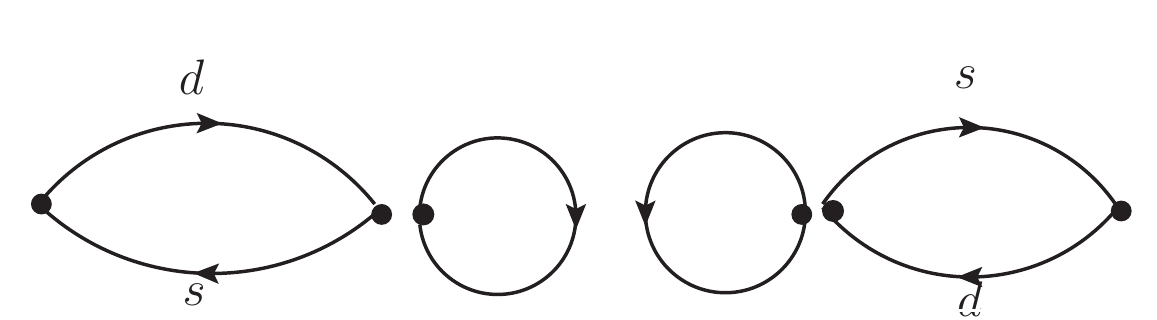}\\Type 4}
\caption{Four types of quark contractions for the calculation of $\Delta M_K$.}
\label{fig:KL_KS_contraction}
\end{figure}

As a subsequent work of Refs.~\cite{Christ:2012se,Bai:2014cva}, a recent calculation of $\Delta M_K$ is performed
on a $2+1$ flavor $32^3\times64$ M\"obius domain wall lattice 
with the Iwasaki + DSDR gauge action.
A near-physical pion mass $M_\pi=170$ MeV and the kaon mass $M_K=492$ MeV are used. Since the calculation is performed
at a coarse lattice spacing with $a^{-1}=1.38$ GeV, the charm quark mass $m_c^{\MS}(\mbox{3 GeV})=750$ MeV 
is unphysically light. The calculation has included all the contractions from Type 1 to Type 4 shown 
in Fig.~\ref{fig:KL_KS_contraction}. Based on 120 configurations, the preliminary lattice 
result is given by $\Delta M_K=3.85(46)\times10^{-12}$ MeV, which is consistent with the experimental 
value $\Delta M_K=3.483(6)\times10^{-12}$ MeV~\cite{Patrignani:2016xqp}.
However, since the calculation uses unphysical kinematics, this agreement could easily be fortuitous.
Note that in the calculation of $\Delta M_K$, the loop integral involves double Glashow-Iliopoulos-Maiani 
(GIM) cancellation~\cite{Glashow:1970gm} and thus, there is no short-distance divergence. 
On the other hand, the double GIM subtraction makes $\Delta M_K$ significantly rely on the charm quark mass. 
As a consequence, it is important to carry out the calculation at the physical charm quark mass.

A new RBC-UKQCD project, reported by C. Sachrajda~\cite{Sachrajda}, uses both physical pion and 
charm quark masses in the calculation. The computation of $\Delta M_K$ is performed on a $64^3\times128$ lattice with the
 Iwasaki gauge action and M\"obius domain wall fermions at an inverse lattice
spacing of 2.359(7)\,GeV. Various techniques such as the use of all-to-all propagators and all mode averaging
 are used to reduce the statistical uncertainty. Based on 59 configurations, the preliminary result 
 of $\Delta M_K=5.5(1.7)\times10^{-12}$ MeV is consistent with the experimental value. The project has planned to collect 160 measurements in total.

The status and prospects of the determination of $\epsilon$ are updated by W. Lee in Refs.~\cite{Lee:2016xkb,Lee}. 
The estimate of $\epsilon$ is made using the FLAG value for $B_K$, the angle-only-fit results for 
the Wolfenstein parameters and the CKM matrix element $V_{cb}$ from exclusive or inclusive decays. The preliminary results for $\epsilon$ yield
\be
|\epsilon|_{\mathrm{SM}}=
\begin{cases}
1.66(17)\times10^{-3}, & \mbox{using exclusive $V_{cb}$ (Lattice QCD)},\\
2.10(21)\times10^{-3}, & \mbox{using inclusive $V_{cb}$ (QCD sum rule)}.
\end{cases}
\ee
Here the exclusive $V_{cb}$ is determined using the experimental measurements of $\bar{B}\to D^*\ell\bar{\nu}$ and $\bar{B}\to D\ell\bar{\nu}$
together with the lattice QCD calculation for the corresponding hadronic matrix 
elements~\cite{Bailey:2014tva,DeTar:2015orc,Amhis:2016xyh,Bigi:2016mdz}. 
The inclusive $V_{cb}$ is determined using the inclusive decay process $\bar{B}\to X_c\ell\bar{\nu}$ 
and QCD sum rules~\cite{Gambino:2016jkc}. 
When using exclusive $V_{cb}$ as input,
there is a 3.3 $\sigma$ deviation between Standard Model value and experimental measurement 
$|\epsilon|_{\mathrm{exp}}=2.228(11)\times10^{-3}$. Besides, 
$V_{cb}$ dominates the current 10\% Standard Model uncertainty for $\epsilon$. 
Therefore, it is important to have an accurate determination of $V_{cb}$. 
On the other hand, it is also important to compute the long-distance contribution 
to $\epsilon$ precisely, whose size is expected to be a few percent but remains not well understood.

To calculate the long-distance contribution to $\epsilon$, it is better 
to write the GIM cancellation by subtracting the charm quark propagator~\cite{Christ:2012se,Bai:2016gzv}
\be
\sum_{q=u,c,t}\frac{\lambda_q{\slashed p}}{p^2+m_q^2}=\lambda_u\left\{\frac{{\slashed p}}{p^2+m_u^2} - \frac{{\slashed p}}{p^2+m_c^2}\right\}
+\lambda_t\left\{\frac{{\slashed p}}{p^2+m_t^2} - \frac{{\slashed p}}{p^2+m_c^2}\right\}.
\ee
By doing so, the double GIM subtraction results in
three terms in the effective Hamiltonian, with the coefficients $\lambda_u^2$, 
$\lambda_u\lambda_t$ and $\lambda_t^2$, respectively. The $\lambda_u^2$ term is irrelevant
for $\epsilon$. The $\lambda_t^2$ term is purely short-distance dominated. Therefore the only 
interesting term for lattice QCD calculation is the $\lambda_u \lambda_t$ term.

\begin{figure}
        \centering
        \shortstack{\includegraphics[width=.21\textwidth]{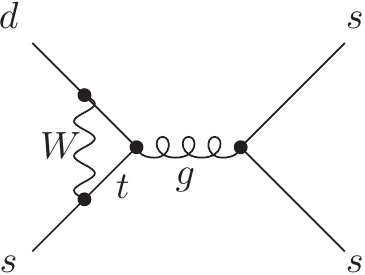}}
        \shortstack{\vspace{0.8cm}\includegraphics[width=.05\textwidth]{arrow}}
        \shortstack{\includegraphics[width=.13\textwidth]{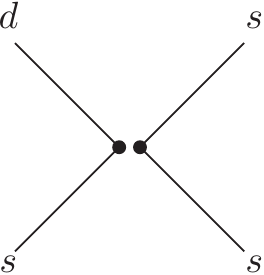}}
        \shortstack{\vspace{0.8cm}\includegraphics[width=.05\textwidth]{arrow}}
        \shortstack{\includegraphics[width=.45\textwidth]{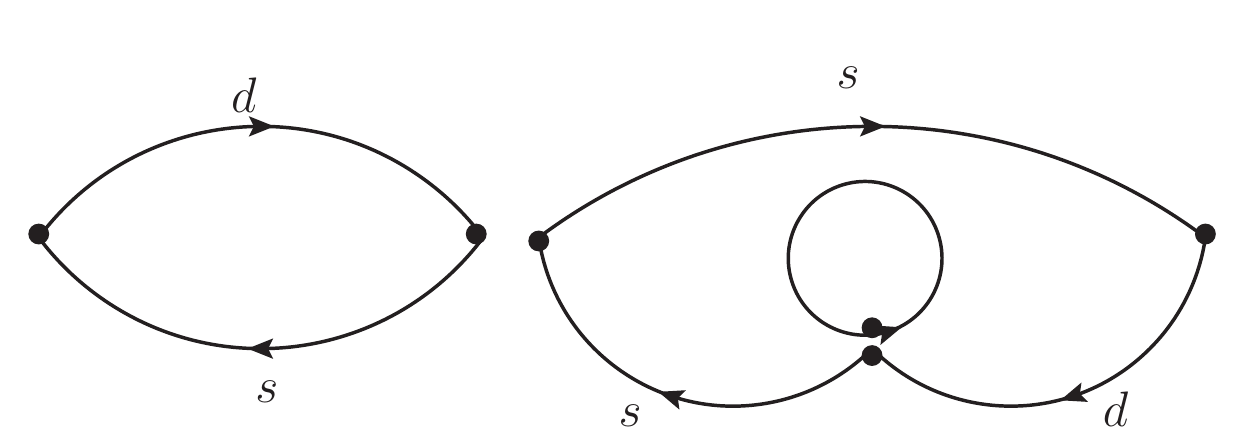}\\Type 5}
		\caption{QCD penguin operator and the origin of the Type 5 diagram for the calculation of the long-distance contribution to $\epsilon$.}
		\label{fig:type5}
    \end{figure}

In the lattice QCD calculation of $\lambda_u \lambda_t$ contribution, 
the top quark field shall be integrated out, leaving a QCD penguin operator, 
shown in Fig.~\ref{fig:type5}. This QCD penguin operator can be neglected 
in the calculation of $\Delta M_K$ as it carries a suppression factor of $\lambda_t/\lambda_u$, 
but it is important for $\epsilon$. The QCD penguin operator together with the current-current operator can form a new Type 5 diagram.

Without top quark in the lattice calculation, there is only one GIM subtraction and 
as a consequence the loop integral is logarithmic divergent. This divergence is 
cut off by an unphysical lattice scale, the inverse lattice spacing $1/a$. 
One can define a bilocal operator in the RI-SMOM scheme by subtracting the 
unphysical short-distance contribution, and then match the bilocal operator in
the RI-SMOM scheme to the one in the $\MS$ scheme using perturbation theory. 
More details on short-distance correction can be found in Refs.~\cite{Christ:2016eae,Bai:2016gzv,Bai:2017fkh}.

The calculation of $\epsilon$ is performed on a $24^3\times64$ lattice with domain wall 
fermion and Iwasaki gauge action~\cite{Bai:2016gzv}. The inverse lattice spacing is $a^{-1}$ is 1.78 GeV. 
The pion mass is 339 MeV and the kaon mass 592 MeV. It uses 200 configurations and 
includes all Type 1-5 diagrams. In Table~\ref{tab:epsilon}
the preliminary lattice results for long-distance contribution to $\epsilon$, $\epsilon^{\mathrm{LD}}$, 
are shown at various RI-SMOM scale $\muRI$ ranging from 1.54 to 2.56 GeV. 
The $\muRI$ dependence is accounted for as a systematic uncertainty. At $\muRI$=2.11 GeV, 
the long-distance contribution to $\epsilon$ is about 5\% when compared to the experimental 
value $|\epsilon|_{\mathrm{exp}}=2.228(11)\times10^{-3}$. To accurately estimate 
the long-distance contribution, the calculation needs to be performed at the physical kinematics.

\begin{table}[ht]
\caption{The long-distance contribution to $\epsilon$ at various $\muRI$, given in units of $10^{-3}$.}
  \centering
  \begin{tabular}{c|c|c|c|c|c}\hline\hline
 $\muRI$ & 1.54 GeV & 1.92 GeV & 2.11 GeV & 2.31 GeV & 2.56 GeV \\ \hline
 $\epsilon^{\mathrm{LD}}$ & $0.091(76)$ & $0.104(76)$ & $0.108(76)$ &  $0.111(77)$ & $0.111(77)$\\
\hline\hline
         \end{tabular}
	\label{tab:epsilon}
 \end{table}

\subsection{Long-distance contributions to flavor changing process: rare kaon decays}

Rare kaon decays have attracted increasing interest during the past few decades. As flavor changing 
neutral current  processes, these decays are highly suppressed in the Standard Model and thus provide 
ideal probes for the observation of new physics effects. In this review, I will discuss the lattice QCD calculations of two classes 
of rare kaon decays: $K\to\pi\nu\bar{\nu}$ and $K\to\pi\ell^+\ell^-$~\cite{Feng:2015kfa,Christ:2015aha,Christ:2016eae,Christ:2016psm,Christ:2016awg,Christ:2016lro,Christ:2016mmq,Bai:2017fkh}.

     \begin{figure}
         \centering
	\shortstack{\includegraphics[width=.3\textwidth]{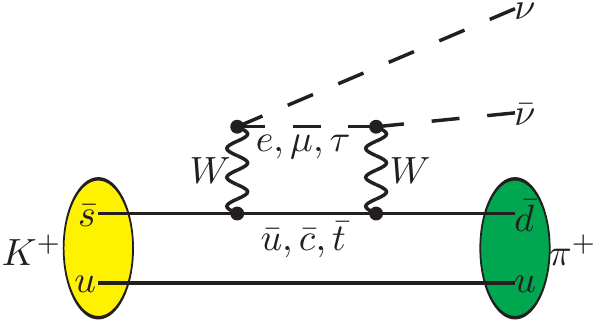}\\$W$-$W$}
	 \hspace{0.5cm}
         \shortstack{\includegraphics[width=.3\textwidth]{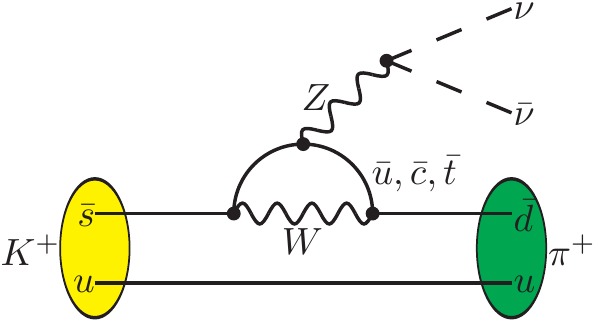}\\$Z$-exchange}
	\caption{Examples of $W$-$W$ and $Z$-exchange diagrams for $K^+\to\pi^+\nu\bar{\nu}$ decay.}
	\label{fig:WW_Z}
     \end{figure}

The $K^+ \to \pi^+ \nu \bar{\nu}$ decay is interesting because it receives the largest contribution 
from top quark loop and thus theoretically very clean. The required hadronic matrix elements 
can be obtained from leading order semi-leptonic $K$ decays, such as $K^+\to\pi^+e\bar{\nu}$, 
via isospin rotation. The remaining long-distance contributions below the charm scale are 
expected to be a few percent. Though small, by including the long-distance contribution 
estimated from Ref.~\cite{Isidori:2005xm}, the branching ratio $\mathrm{Br}(K^+\to\pi^+\nu\bar{\nu})$ 
is enhanced by 6\%, which is comparable to the 8\% total Standard Model uncertainty~\cite{Buras:2015qea}. 
The current known branching-ratio measurement~\cite{Artamonov:2008qb}
\be
\mathrm{Br}(K^+\to\pi^+\nu\bar{\nu})_{\mathrm{exp}}=1.73^{+1.15}_{-1.05}\times10^{-10}
\ee
 is a combined result based on the 7 events collected by BNL 
 E787~\cite{Adler:1997am,Adler:2000by,Adler:2001xv,Adler:2002hy} and its 
 successor E949~\cite{Anisimovsky:2004hr,Artamonov:2008qb}. Its central value is 
 almost twice of the Standard Model prediction~\cite{Buras:2015qea}
\be
\mathrm{Br}(K^+\to\pi^+\nu\bar{\nu})_{\mathrm{SM}}=9.11\pm0.72\times10^{-11},
\ee 
but with a 60-70\% uncertainty it is still consistent with Standard Model.

The new experiment, NA62 in CERN~\cite{fortheNA62:2013jsa}, aims at an observation of $O(100)$ 
events and a 10\%-precision measurement of $\mathrm{Br}(K^+\to\pi^+\nu\bar{\nu})$. 
The status reported at the Flavor physics and CP violation workshop (FPCP 2017)
is that the detector installation is completed in September 2016. 5\% of the 2016 
data has been analyzed but no event is found yet. If using full 2016 data, then 
$O(1)$ events are expected to be found. 
Considering the fact that the Standard Model predictions will be confronted with the new experiment soon, 
a lattice QCD calculation of the long-distance contribution to $K^+\to\pi^+\nu\bar{\nu}$ is timely.

There are two classes of diagrams, which contribute to $K^+\to\pi^+\nu\bar{\nu}$ decays, called as $W$-$W$ 
and $Z$-exchange diagrams. In the $W$-$W$ diagrams the second-order weak transition proceeds through 
the exchange of two $W$-bosons, while for the $Z$-exchange diagrams
the decay occurs through the exchange of one $W$-boson and one $Z$-boson. Examples of both classes 
of diagrams are illustrated in Fig.~\ref{fig:WW_Z}.

In a lattice QCD calculation, the $W$ and $Z$-boson have been integrated out, leaving two effective 
four-fermion local operators. The matrix element of the time-integrated bilocal operator is 
evaluated in Euclidean space.  This matrix element can be related to the second-order amplitude 
of interest if a sum over intermediate states is inserted and the integration over Euclidean time performed:
\ba
&&\int_{-T}^{T} dt\, \langle \pi^+\nu\overline{\nu}| T\left\{H_A(t)H_B(0)\right\}|K^+\rangle 
\nn\\
&&\hspace{1.5cm}=  \sum_n \Biggl\{ \frac{\langle \pi^+\nu\overline{\nu}|H_A|n\rangle\langle
    n|H_B|K^+\rangle}{E_n-E_K}+\frac{\langle
    \pi^+\nu\overline{\nu}|H_B|n\rangle\langle
n|H_A|K^+\rangle}{E_n-E_K}\Biggr\}
                           \left(1-e^{(E_K-E_n)T}\right),
\ea
where $H_{A/B}(t)$ stands for the two four-fermion operators, with the spatial variables integrated over space.
 The unphysical $e^{(E_K-E_n)T}$ terms in the second line of this equation vanish for large $T$ for 
 intermediate states more energetic than the kaon.  However, these terms grow exponentially with 
 increasing integration range if $E_n < E_K$ and must be removed from lattice calculation.
When the intermediate state involves multiple particles, the branch-cut integral in the infinite 
volume is replaced by a discrete state summation in the finite volume. It could cause potentially 
large finite-volume effects when $E_n\to M_K$, which need to be corrected following Ref.~\cite{Christ:2015pwa}.

\begin{figure}[tp]
   \centering
             {\includegraphics[width=0.7\textwidth,clip]{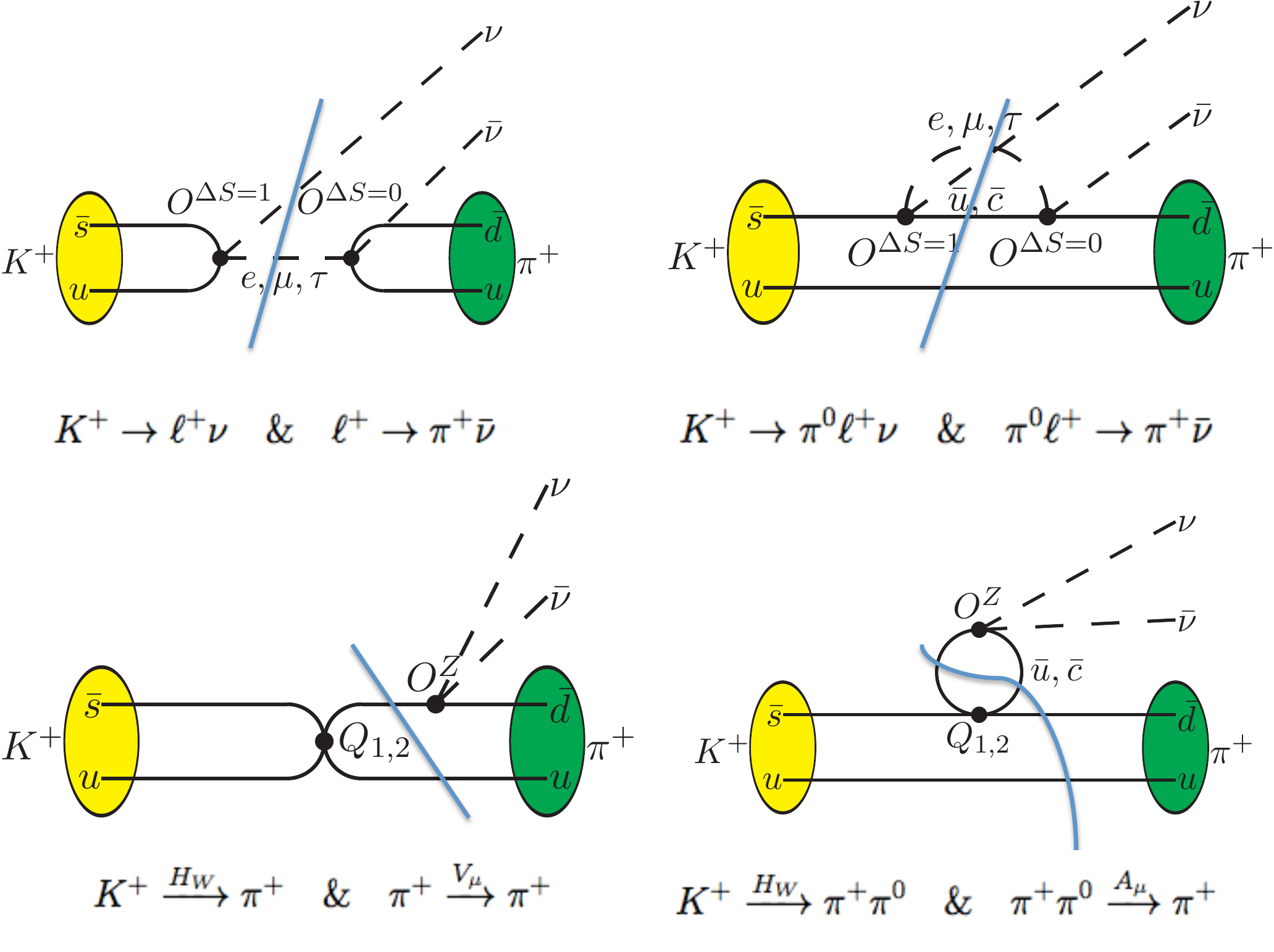}}
	\caption{Low-lying intermediate states contributing to $K^+\to\pi^+\nu\bar{\nu}$. As these states 
    are related to exponentially growing unphysical contributions and potentially large finite volume effects, 
one shall calculate the hadronic matrix elements for these low-lying intermediate states from the relevant 2-point and 3-point functions.}
	\label{fig:low-lying-state}
\end{figure}

To deal with the exponentially growing terms and finite volume effects, the matrix elements for the 
lowing-lying intermediate states shall be calculated. These states include the leptonic $\ell^+\nu$, 
semileptonic $\pi^0\ell^+\nu$, single pion and isospin $I=2$ $\pi^+\pi^0$ scattering state and are 
summarized by Fig.~\ref{fig:low-lying-state}. So the study of the long-distance contribution to 
$K^+\to\pi^+\nu\bar{\nu}$ decay does not only involve the calculation of 4-point function, 
but also includes the calculation of all relevant 2-point and 3-point functions for low-lying intermediate states.

\begin{figure}
\centering
\includegraphics[width=.8\textwidth]{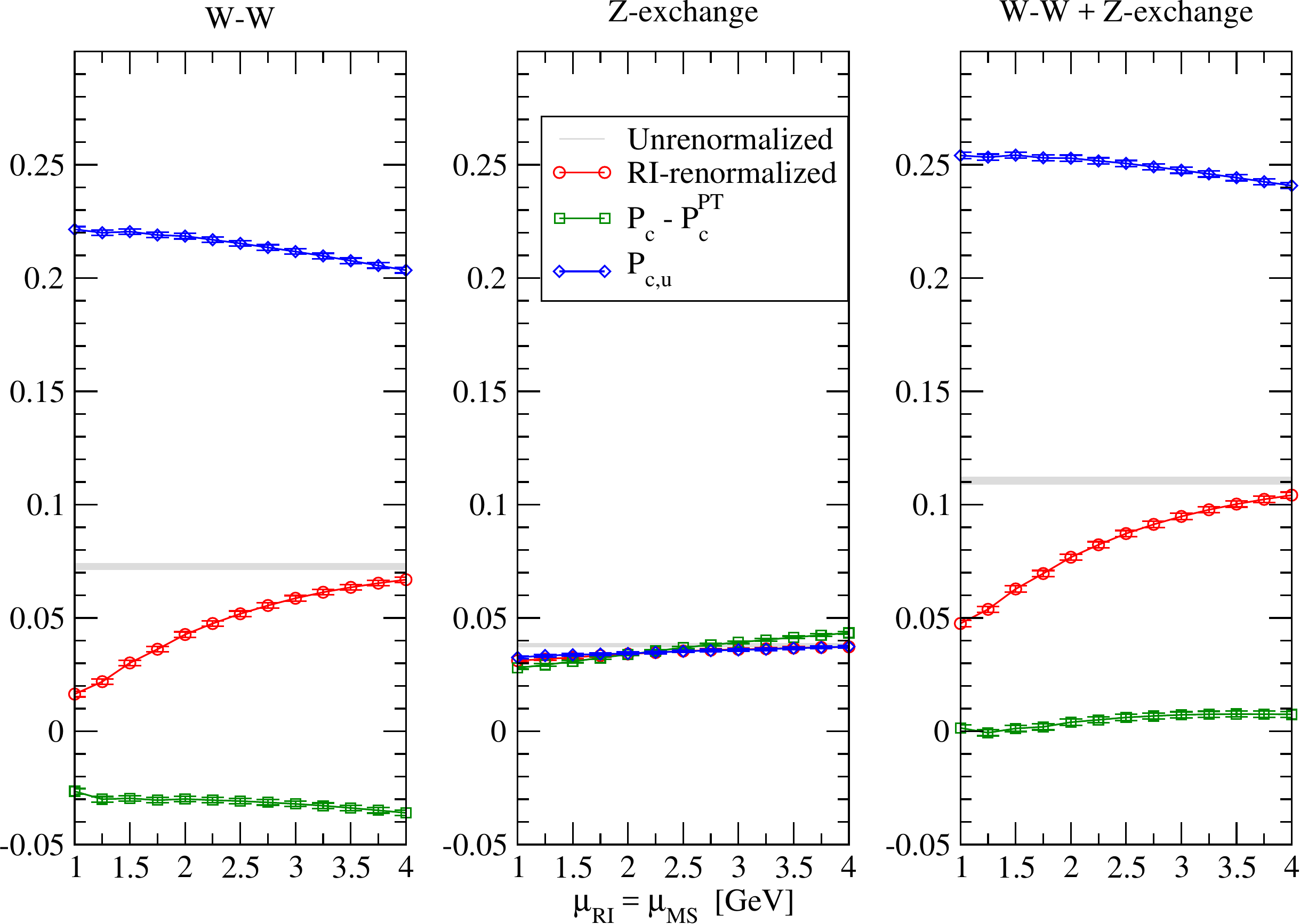}
\caption{$W$-$W$ and $Z$-exchange diagram results, and their total, shown from left to the right. 
The gray bands show the bilocal matrix element including the unphysical lattice artifacts. 
The red circles indicate the RI-renormalized, bilocal contribution. The blue diamonds give the 
total charm contribution $P_c$ while the green squares show the difference between the lattice and perturbative results, $P_c - P^{\mathrm{PT}}_c$. }
\label{fig:Pc}
\end{figure}

As the top quark contribution to the decay is completely short-distance dominated, 
one only needs to focus on the charm quark contribution.
The first calculation is performed using the $16^3\times32$, $N_f=2+1$ flavor, 
domain wall fermion ensemble, with $a^{-1}=1.729(28)$ GeV~\cite{Bai:2017fkh}. This 
ensemble has pion and kaon masses of $M_\pi\sim421$ MeV and $M_K\sim563$ MeV. 
The $\MS$ charm quark mass is $m_c^{\MS}(\mbox{2 GeV})\sim863$ MeV. Both $W$-$W$ 
and $Z$-exchange diagrams are logarithmically divergent and cutoff by unphysical 
scale $1/a$. Similar as the computation of $\epsilon$, the short-distance correction 
needs to be performed here~\cite{Christ:2016eae}.
The lattice results are shown in Fig.~\ref{fig:Pc}. Here $P_c$ gives the complete 
charm quark contribution to the $K^+\to\pi^+\nu\bar{\nu}$ decay.
The results from the $W$-$W$ and $Z$-exchange diagrams, and their total, 
are shown in the left, center and
right panels. The gray bands show the bilocal matrix element including the unphysical 
lattice artifacts. The red circles indicate the RI-renormalized, bilocal contribution. 
The blue diamonds give the total charm
contribution $P_c$, while the green squares show the difference
between the lattice and perturbative results, $P_c-P_c^{\mathrm{PT}}$. 
The results from the exploratory lattice calculation with unphysical charm, down and up quark masses are:
\be
P_c=0.2529(\pm13)(\pm32)(-45),
\quad P_c-P_c^{\mathrm{PT}}=0.0040(\pm13)(\pm32)(-45),
\ee
The small size of $P_c-P_c^{\mathrm{PT}}$ results from a large cancellation 
between the $W$-$W$ and $Z$-exchange amplitudes. It is important to determine 
whether such a large cancellation persists for physical quark masses.

Different from $K^+\to\pi^+\nu\bar{\nu}$ decay, 
the CP conserving decays $K^+\to\pi^+\ell^+\ell^-$ and $K_S\to\pi^0\ell^+\ell^-$ 
receive the dominated long-distance contribution from $\gamma$-exchange diagram.
 Although the loop integral in the $\gamma$-exchange diagram is quadratically ultraviolet divergent by power counting,
the electromagnetic gauge invariance reduces the divergence to be logarithmic.
The GIM cancellation further reduces the logarithmic divergence to be ultraviolet finite.

\begin{figure}
\centering
\includegraphics[width=.8\textwidth]{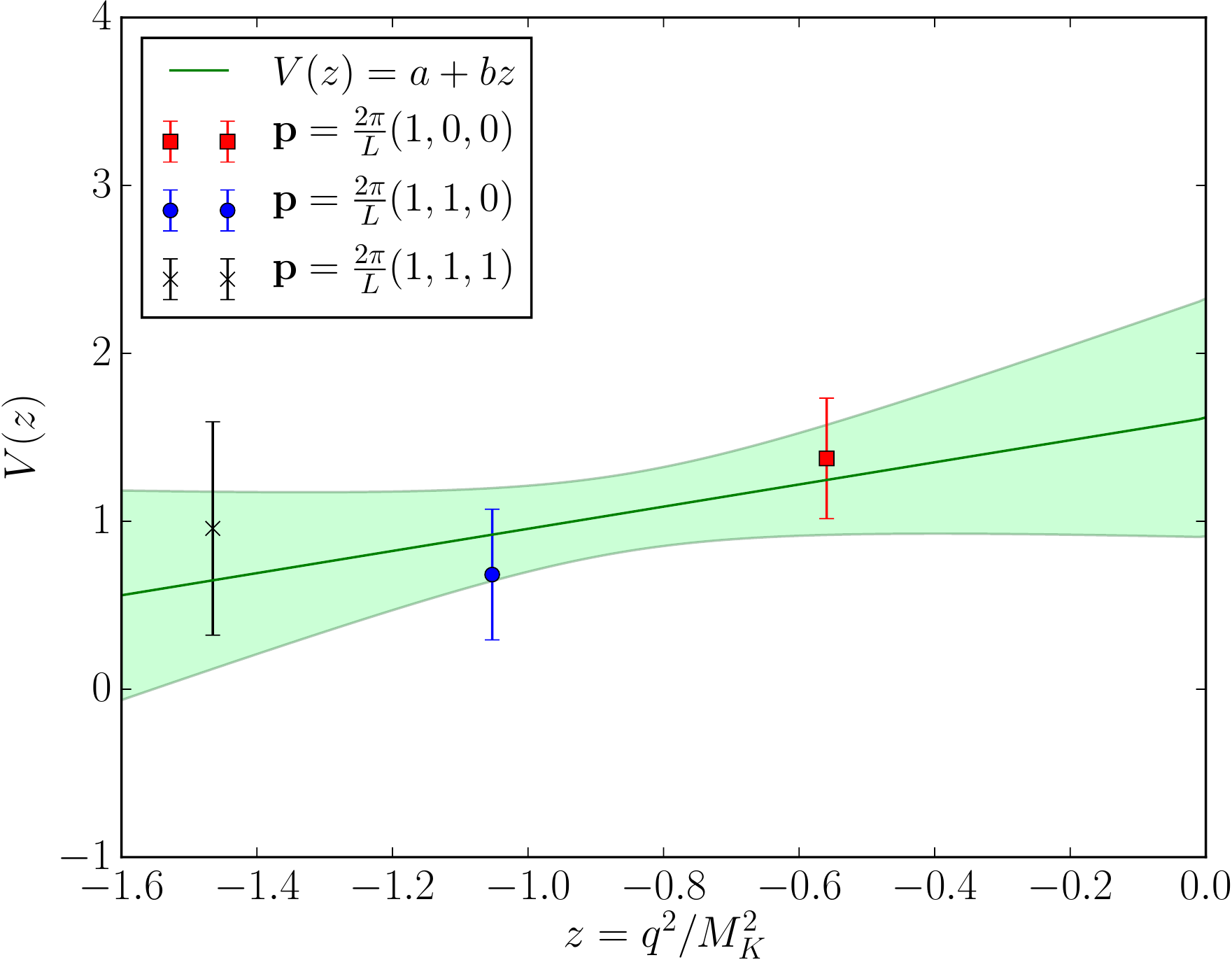}
\caption{Dependence of the form factor for the decay $K^+\to\pi^+\ell^+\ell^-$ upon $z=q^2/M_K^2$. The lattice data is fit to a linear form
$V_+(z)=a_++b_+z.$}
\label{fig:V_+}
\end{figure}

In the $\gamma$-exchange process, the hadronic part of amplitudes for $K^+$ and $K_S$ decays 
can be written in terms of electromagnetic transition form factor $V_{+/S}(z)$ via~\cite{Cirigliano:2011ny}
\ba
T_{+/S}^\mu(p_K,p_\pi)&=&\int d^4x\,e^{iqx}\langle\pi^{+/0}(p_\pi)|T\{J_{em}^\mu(x)\mathcal{H}^{\Delta S=1}(0)\}|K^+/K_S(p_K)\rangle.
\nn\\
&=&\frac{G_FM_K^2}{(4\pi)^2}V_{+/S}(z)\left[z(p_K+p_\pi)^\mu-(1-r_\pi^2)q^\mu\right]
\ea
with $p_{K/\pi}$ the kaon/pion momentum, $q=p_K-p_\pi$, $z=q^2/M_K^2$ and $r_\pi=M_\pi/M_K$.
The target for lattice QCD calculation is to extract $V_{+/S}(z)$ from the bilocal hadronic matrix elements by building the relevant 4-point correlation functions.
The strategy adopted in Ref.~\cite{Christ:2015aha} is to use conserved vector current to protect 
the electromagnetic gauge invariance and use the charm quark as an active quark flavor to maintain GIM cancellation.
The first exploratory calculation of $K^+\to\pi^+\ell^+\ell^-$ decay~\cite{Christ:2016mmq} is performed using a $24^3\times64$ lattice with
domain wall fermion and Iwasaki gauge action.
The inverse lattice spacing is $1/a=1.78$ GeV. The calculation uses
a pion mass of $M_\pi\sim430$ MeV, a kaon mass of $M_K\sim$ 625 MeV and a $\MS$ charm quark mass 
$m_c^{\MS}(\mbox{2 GeV})\sim533$ MeV. Three momentum transfers are used in the calculation and 
a linear fit form $V_+(z)=a_++b_+z$ is used to determine the momentum dependence of the form factor. 
The lattice data points for $V_+(z)$ together with the fit curve are shown in Fig.~\ref{fig:V_+}. 
Using 128 configurations, the results for $a_+$ and $b_+$ yield
\be
\label{eq:lattice_result}
a_+=1.6(7),\quad b_+=0.7(8).
\ee

The phenomenological study~\cite{DAmbrosio:1998gur} decomposes the form factor into a linear 
form plus the unitarity $\pi\pi$-loop correction $V_+^{\pi\pi}(z)$
\be
\label{eq:V_+_fit}
V_+(z)=a_++b_+z+V_+^{\pi\pi}(z).
\ee
Here $V_+^{\pi\pi}(z)$ is determined using chiral perturbation theory together with 
some model assumptions such as vector meson dominance model. 
The experimental measurements of the branching ratio together with the fit form~(\ref{eq:V_+_fit}) 
produce the much more precise results for $a_+$ and $b_+$~\cite{Batley:2009aa,Batley:2011zz}
\be
\label{eq:pheno_result}
\begin{cases}
a_+=-0.58(2),\quad b_+=-0.78(7), & \mbox{fit from $K^+\to\pi^+e^+e^-$ spectra},\\
a_+=-0.58(4),\quad b_+=-0.81(15), & \mbox{fit from $K^+\to\pi^+\mu^+\mu^-$ spectra},
\end{cases}
\ee
where the first line uses the experimental measurement of $K^+\to\pi^+e^+e^-$ spectra and the second line uses $K^+\to\pi^+\mu^+\mu^-$ data. 
Note that these results carry the opposite signs to the lattice results in Eq.~(\ref{eq:lattice_result}).
Since the lattice calculation is performed at unphysical quark masses, it does not make much 
sense to compare these results in Eqs.~(\ref{eq:lattice_result}) and (\ref{eq:pheno_result}). 
On the other hand,  
as the experimental data only yield the square of the form factor and does not tell the sign for $V_+(z)$, 
the signs for $a_+$ and $b_+$ are completely determined by the input of $V_+^{\pi\pi}(z)$. However,
it is found that the polynomial contribution (linear in $z$) dominates over 
the unitarity loop correction. For $K^+\to\pi^+\mu^+\mu^-$ decay, the fit forms with and without $V_+^{\pi\pi}$ 
correction produce almost the same fit curves. For $K^+\to\pi^+e^+e^-$ decay, the fit curves differ at small $z$, 
where the experimental data is not available~\cite{Cirigliano:2011ny}. Therefore it is questionable to use 
$V_+^{\pi\pi}(z)$ to determine the signs for $a_+$ and $b_+$.
It is important to perform a lattice QCD calculation at the physical quark mass and examine the 
phenomenological fit ansatz~(\ref{eq:V_+_fit}) and confirm the sign for $a_+$ and $b_+$.

In order to perform the calculation at the physical point, the physical pion mass would require the
large lattice volume and physical charm quark mass 
would require for the ultra-fine lattice spacing to control both finite-volume effects and lattice artifacts. 
Thus it is very high demanding on computer resources.
One solution is to improve quark action to reduce the lattice artifacts for the charm quark. 
In Ref.~\cite{Tomii:2017lyo} M. Tomii performs an exploratory study of dispersion relation and 
unphysical pole for Mobius domain wall fermion and seeks for a way to improve the action.
Another solution is to integrate out of charm quark field using perturbation theory. In this case, 
lattice QCD calculation only requires the physical pion mass and a rather coarse lattice spacing. 
This would save the computer resources quite significantly. But a drawback is that 
since there is no GIM cancellation, the internal up quark loop will be logarithmically divergent. 
In Ref.~\cite{Lawson} A. Lawson discusses on the renormalization to treat with
the short-distance divergence in three-flavor theory.

Besides for the CP conserving $K \to \pi\ell^+\ell^-$ decay, it is
also interesting to study the CP violating $K_L$ decays. 
The $K_L$ decay amplitudes receive three major contributions:
1) a short-distance dominated direct CP violation, 2) a long-distance dominated, indirect CP violating contribution through
$K_L\to K_+\to\pi^0\ell^+\ell^-$, 3) a CP conserving component which proceeds through two-photon exchange.
Total CP violating contributions to $K_L$ decay branching ratios, including 1), 2) and their interference, are given by~\cite{Buchalla:2003sj,Isidori:2004rb}
\ba
&&\mathrm{Br}(K_L\to\pi^0e^+e^-)_{\mathrm{CPV}}=10^{-12}\times\left[15.7|a_S|^2\pm6.2|a_S|\left(\frac{\operatorname{Im}\lambda_t}{10^{-4}}\right)
+2.4\left(\frac{\operatorname{Im}\lambda_t}{10^{-4}}\right)\right],
\nn\\
&&\mathrm{Br}(K_L\to\pi^0\mu^+\mu^-)_{\mathrm{CPV}}=10^{-12}\times\left[3.7|a_S|^2\pm1.6|a_S|\left(\frac{\operatorname{Im}\lambda_t}{10^{-4}}\right)
+1.0\left(\frac{\operatorname{Im}\lambda_t}{10^{-4}}\right)\right],
\ea
where the $\lambda_t\approx1.35\times10^{-4}$. The parameter $a_S$ is given by the $K_S$ transition form factor at zero momentum transfer, namely
 $a_S=V_S(0)$, and it is a quantity of size $O(1)$. The $\pm$ sign arises because only the magnitude of $a_S$ is determined from experiment.
Therefore even a determination of the sign of $a_S$ from lattice QCD is desirable.

\section{Conclusion}

The worldwide lattice QCD community has developed a successful kaon physics program. It inspires the consideration of constructing
a CKM unitarity triangle purely from kaon physics~\cite{Lehner:2015jga}.

 For standard quantities such as $f_{K^\pm}/f_{\pi^\pm}$, $f_+(0)$ and $\hat{B}_K$, they are computed with a 
 precision of $\sim$ 1 percent or even much better, as shown in Table~\ref{tab:FLAG}. In these cases lattice 
 QCD calculations play important roles in precision flavor physics.
With the development of the lattice QCD techniques, it is also the time to
explore the non-standard quantities. Here I report the recent progress of 
the calculations on $K\to\pi\pi$ decay, long-distance contributions to $\Delta M_K$ 
and $\epsilon$ as well as rare kaon decays. Lattice QCD is now capable of first-principals 
calculation of these non-standard quantities. Some of the calculations are even performed at the physical kinematics.
We can foresee that with the new techniques and new generation of super-computers 
today's non-standard observables will become standard in the near future. 

     \begin{table}[ht]
	\caption{Summary of the FLAG average of $f_{K^\pm}/f_{\pi^\pm}$, $f_+(0)$ and $\hat{B}_K$.}
         \centering
         \begin{tabular}{cccc}\hline
             & $N_f$ & FLAG average & Frac. Err. \\ \hline
             $f_{K^\pm}/f_{\pi^\pm}$ & $2+1+1$ & 1.1933(29) & 0.25\% \\
             $f_+(0)$ & $2+1+1$ & 0.9706(27) & 0.28\% \\
             $\hat{B}_K$ & $2+1$ & 0.7625(97) & 1.27\% \\
             \hline
         \end{tabular}
	\label{tab:FLAG}
     \end{table}

\vspace{0.3cm}

Acknowledgments. -- I warmly thank my colleagues from the RBC-UKQCD
collaboration for many helpful discussions. I also gratefully acknowledge the
material supply from Z. Bai, N. H. Christ, E. G\'amiz, N. Garron, T. Izubuchi,
T. Kaneko, C. Kelly, J. Kettle, A. Lawson, W. Lee, H. Ohki, C. T. Sachrajda, 
and A. Soni when I made the preparation for this review.
This work was supported in part by the National Natural Science Foundation of
China under Grant No. 11775002.

\clearpage
\bibliography{lattice2017}

\begin{thebibliography}{76}

\bibitem{Aoki:2016frl}
S.~Aoki et~al., Eur. Phys. J. \textbf{C77}, 112 (2017), \texttt{1607.00299}

\bibitem{Moulson:2014cra}
M.~Moulson, \emph{{Experimental determination of $V_{us}$ from kaon decays}},
  in \emph{{8th International Workshop on the CKM Unitarity Triangle (CKM 2014)
  Vienna, Austria, September 8-12, 2014}} (2014), \texttt{1411.5252},
  \urlstyle{tt}\url{https://inspirehep.net/record/1328784/files/arXiv:1411.5252.pdf}

\bibitem{Rosner:2015wva}
J.L. Rosner, S.~Stone, R.S. Van~de Water, Submitted to: Particle Data Book
  (2015), \texttt{1509.02220}

\bibitem{Hardy:2014qxa}
J.C. Hardy, I.S. Towner, Phys. Rev. \textbf{C91}, 025501 (2015),
  \texttt{1411.5987}

\bibitem{Hardy:2016vhg}
J.~Hardy, I.S. Towner, PoS \textbf{CKM2016}, 028 (2016)

\bibitem{Gamiz:2016bpm}
E.~Gámiz et~al. (Fermilab Lattice, MILC), PoS \textbf{LATTICE2016}, 286
  (2016), \texttt{1611.04118}

\bibitem{Bernard:2017scg}
C.~Bernard, J.~Bijnens, E.~Gámiz, J.~Relefors, JHEP \textbf{03}, 120 (2017),
  \texttt{1702.03416}

\bibitem{Aoki:2015pba}
S.~Aoki, G.~Cossu, X.~Feng, S.~Hashimoto, T.~Kaneko, J.~Noaki, T.~Onogi
  (JLQCD), Phys. Rev. \textbf{D93}, 034504 (2016), \texttt{1510.06470}

\bibitem{Aoki:2017spo}
S.~Aoki, G.~Cossu, X.~Feng, H.~Fukaya, S.~Hashimoto, T.~Kaneko, J.~Noaki,
  T.~Onogi (JLQCD), Phys. Rev. \textbf{D96}, 034501 (2017), \texttt{1705.00884}

\bibitem{Amhis:2016xyh}
Y.~Amhis et~al. (2016), \texttt{1612.07233}

\bibitem{Patrignani:2016xqp}
C.~Patrignani et~al. (Particle Data Group), Chin. Phys. \textbf{C40}, 100001
  (2016)

\bibitem{Braaten:1991qm}
E.~Braaten, S.~Narison, A.~Pich, Nucl. Phys. \textbf{B373}, 581 (1992)

\bibitem{Erler:2002mv}
J.~Erler, Rev. Mex. Fis. \textbf{50}, 200 (2004), \texttt{hep-ph/0211345}

\bibitem{Gamiz:2004ar}
E.~Gamiz, M.~Jamin, A.~Pich, J.~Prades, F.~Schwab, Phys. Rev. Lett.
  \textbf{94}, 011803 (2005), \texttt{hep-ph/0408044}

\bibitem{Hudspith:2017vew}
R.J. Hudspith, R.~Lewis, K.~Maltman, J.~Zanotti (2017), \texttt{1702.01767}

\bibitem{Ohki}
H.~Ohki et~al., \emph{{$|V_{us}|$ determination from inclusive strange tau
  decay and lattice HVP}}, in \emph{Proceedings,
  \href{http://inspirehep.net/record/1425631}{35th International Symposium on
  Lattice Field Theory (Lattice2017)}: Granada, Spain}, to appear in EPJ Web
  Conf.

\bibitem{Garron:2016mva}
N.~Garron, R.J. Hudspith, A.T. Lytle (RBC/UKQCD), JHEP \textbf{11}, 001 (2016),
  \texttt{1609.03334}

\bibitem{Boyle:2017skn}
P.A. Boyle, N.~Garron, R.J. Hudspith, C.~Lehner, A.T. Lytle (2017),
  \texttt{1708.03552}

\bibitem{Garron}
N.~Garron, \emph{{Fierz relations and renormalization schemes for four-quark
  operators}}, in \emph{Proceedings,
  \href{http://inspirehep.net/record/1425631}{35th International Symposium on
  Lattice Field Theory (Lattice2017)}: Granada, Spain}, to appear in EPJ Web
  Conf.

\bibitem{Aoki:2007xm}
Y.~Aoki et~al., Phys. Rev. \textbf{D78}, 054510 (2008), \texttt{0712.1061}

\bibitem{Boyle:2017ssm}
P.~Boyle, N.~Garron, J.~Kettle, A.~Khamseh, J.T. Tsang (2017),
  \texttt{1710.09176}

\bibitem{Buchalla:1995vs}
G.~Buchalla, A.J. Buras, M.E. Lautenbacher, Rev. Mod. Phys. \textbf{68}, 1125
  (1996), \texttt{hep-ph/9512380}

\bibitem{Blum:2015ywa}
T.~Blum et~al., Phys. Rev. \textbf{D91}, 074502 (2015), \texttt{1502.00263}

\bibitem{Luscher:1990ux}
M.~Luscher, Nucl. Phys. \textbf{B354}, 531 (1991)

\bibitem{Schenk:1991xe}
A.~Schenk, Nucl. Phys. \textbf{B363}, 97 (1991)

\bibitem{GellMann:1955jx}
M.~Gell-Mann, A.~Pais, Phys. Rev. \textbf{97}, 1387 (1955)

\bibitem{Boyle:2012ys}
P.A. Boyle et~al. (RBC, UKQCD), Phys. Rev. Lett. \textbf{110}, 152001 (2013),
  \texttt{1212.1474}

\bibitem{Donini:2016lwz}
A.~Donini, P.~Hernández, C.~Pena, F.~Romero-López, Phys. Rev. \textbf{D94},
  114511 (2016), \texttt{1607.03262}

\bibitem{Bai:2015nea}
Z.~Bai et~al. (RBC, UKQCD), Phys. Rev. Lett. \textbf{115}, 212001 (2015),
  \texttt{1505.07863}

\bibitem{Colangelo:2001df}
G.~Colangelo, J.~Gasser, H.~Leutwyler, Nucl. Phys. \textbf{B603}, 125 (2001),
  \texttt{hep-ph/0103088}

\bibitem{Colangelo:2015kha}
G.~Colangelo, E.~Passemar, P.~Stoffer, Eur. Phys. J. \textbf{C75}, 172 (2015),
  \texttt{1501.05627}

\bibitem{PDG2014}
K.~Olive, P.D. Group, Chinese Physics C \textbf{38}, 090001 (2014)

\bibitem{Kelly}
C.~Kelly et~al., \emph{{Progress in the improved lattice calculation of direct
  CP-violation in the Standard Model}}, in \emph{Proceedings,
  \href{http://inspirehep.net/record/1425631}{35th International Symposium on
  Lattice Field Theory (Lattice2017)}: Granada, Spain}, to appear in EPJ Web
  Conf.

\bibitem{Christ}
N.H. Christ, X.~Feng, \emph{{Including electromagnetism in $K\to\pi\pi$ decay
  calculations}}, in \emph{Proceedings,
  \href{http://inspirehep.net/record/1425631}{35th International Symposium on
  Lattice Field Theory (Lattice2017)}: Granada, Spain}, to appear in EPJ Web
  Conf.

\bibitem{Bruno}
M.~Bruno et~al., \emph{{Weak hamiltonian Wilson Coefficients from Lattice
  QCD}}, in \emph{Proceedings, \href{http://inspirehep.net/record/1425631}{35th
  International Symposium on Lattice Field Theory (Lattice2017)}: Granada,
  Spain}, to appear in EPJ Web Conf.

\bibitem{Ishizuka:2015oja}
N.~Ishizuka, K.I. Ishikawa, A.~Ukawa, T.~Yoshié, Phys. Rev. \textbf{D92},
  074503 (2015), \texttt{1505.05289}

\bibitem{Mohapatra:1968zz}
R.N. Mohapatra, J.S. Rao, R.E. Marshak, Phys. Rev. \textbf{171}, 1502 (1968)

\bibitem{Glashow:1970gm}
S.L. Glashow, J.~Iliopoulos, L.~Maiani, Phys. Rev. \textbf{D2}, 1285 (1970)

\bibitem{Gaillard:1974hs}
M.K. Gaillard, B.W. Lee, Phys. Rev. \textbf{D10}, 897 (1974)

\bibitem{Bai:2016gzv}
Z.~Bai, PoS \textbf{LATTICE2016}, 309 (2017), \texttt{1611.06601}

\bibitem{Christ:2012se}
N.H. Christ, T.~Izubuchi, C.T. Sachrajda, A.~Soni, J.~Yu (RBC, UKQCD), Phys.
  Rev. \textbf{D88}, 014508 (2013), \texttt{1212.5931}

\bibitem{Bai:2014cva}
Z.~Bai, N.H. Christ, T.~Izubuchi, C.T. Sachrajda, A.~Soni, J.~Yu, Phys. Rev.
  Lett. \textbf{113}, 112003 (2014), \texttt{1406.0916}

\bibitem{Sachrajda}
C.T. Sachrajda et~al., \emph{{The $K_L$-$K_S$ Mass Difference}}, in
  \emph{Proceedings, \href{http://inspirehep.net/record/1425631}{35th
  International Symposium on Lattice Field Theory (Lattice2017)}: Granada,
  Spain}, to appear in EPJ Web Conf.

\bibitem{Lee:2016xkb}
W.~Lee, J. Phys. Conf. Ser. \textbf{800}, 012006 (2017), \texttt{1611.04261}

\bibitem{Lee}
W.~Lee, \emph{{Update on $\varepsilon_K$ with lattice QCD inputs}}, in
  \emph{Proceedings, \href{http://inspirehep.net/record/1425631}{35th
  International Symposium on Lattice Field Theory (Lattice2017)}: Granada,
  Spain}, to appear in EPJ Web Conf.

\bibitem{Bailey:2014tva}
J.A. Bailey et~al. (Fermilab Lattice, MILC), Phys. Rev. \textbf{D89}, 114504
  (2014), \texttt{1403.0635}

\bibitem{DeTar:2015orc}
C.~DeTar, PoS \textbf{LeptonPhoton2015}, 023 (2016), \texttt{1511.06884}

\bibitem{Bigi:2016mdz}
D.~Bigi, P.~Gambino, Phys. Rev. \textbf{D94}, 094008 (2016),
  \texttt{1606.08030}

\bibitem{Gambino:2016jkc}
P.~Gambino, K.J. Healey, S.~Turczyk, Phys. Lett. \textbf{B763}, 60 (2016),
  \texttt{1606.06174}

\bibitem{Christ:2016eae}
N.H. Christ, X.~Feng, A.~Portelli, C.T. Sachrajda (RBC, UKQCD), Phys. Rev.
  \textbf{D93}, 114517 (2016), \texttt{1605.04442}

\bibitem{Bai:2017fkh}
Z.~Bai, N.H. Christ, X.~Feng, A.~Lawson, A.~Portelli, C.T. Sachrajda, Phys.
  Rev. Lett. \textbf{118}, 252001 (2017), \texttt{1701.02858}

\bibitem{Feng:2015kfa}
X.~Feng, N.H. Christ, A.~Portelli, C.~Sachrajda, PoS \textbf{LATTICE2014}, 367
  (2015)

\bibitem{Christ:2015aha}
N.H. Christ, X.~Feng, A.~Portelli, C.T. Sachrajda (RBC, UKQCD), Phys. Rev.
  \textbf{D92}, 094512 (2015), \texttt{1507.03094}

\bibitem{Christ:2016psm}
N.H. Christ, X.~Feng, A.~Jüttner, A.~Lawson, A.~Portelli, C.T. Sachrajda, PoS
  \textbf{CD15}, 033 (2016)

\bibitem{Christ:2016awg}
N.~Christ, X.~Feng, A.~Juttner, A.~Lawson, A.~Portelli, C.~Sachrajda,
  \emph{{Long distance contributions to the rare kaon decay
  $K\to\pi\ell^{+}\ell^{-}$}}, in \emph{{Proceedings, 33rd International
  Symposium on Lattice Field Theory (Lattice 2015)}} (2016),
  \texttt{1602.01374},
  \urlstyle{tt}\url{http://inspirehep.net/record/1419253/files/arXiv:1602.01374.pdf}

\bibitem{Christ:2016lro}
N.H. Christ, X.~Feng, A.~Lawson, A.~Portelli, C.~Sachrajda, PoS
  \textbf{LATTICE2016}, 306 (2016)

\bibitem{Christ:2016mmq}
N.H. Christ, X.~Feng, A.~Juttner, A.~Lawson, A.~Portelli, C.T. Sachrajda, Phys.
  Rev. \textbf{D94}, 114516 (2016), \texttt{1608.07585}

\bibitem{Isidori:2005xm}
G.~Isidori, F.~Mescia, C.~Smith, Nucl.Phys. \textbf{B718}, 319 (2005),
  \texttt{hep-ph/0503107}

\bibitem{Buras:2015qea}
A.J. Buras, D.~Buttazzo, J.~Girrbach-Noe, R.~Knegjens (2015),
  \texttt{1503.02693}

\bibitem{Artamonov:2008qb}
A.~Artamonov et~al. (E949 Collaboration), Phys.Rev.Lett. \textbf{101}, 191802
  (2008), \texttt{0808.2459}

\bibitem{Adler:1997am}
S.~Adler et~al. (E787 Collaboration), Phys.Rev.Lett. \textbf{79}, 2204 (1997),
  \texttt{hep-ex/9708031}

\bibitem{Adler:2000by}
S.~Adler et~al. (E787 Collaboration), Phys.Rev.Lett. \textbf{84}, 3768 (2000),
  \texttt{hep-ex/0002015}

\bibitem{Adler:2001xv}
S.~Adler et~al. (E787 Collaboration), Phys.Rev.Lett. \textbf{88}, 041803
  (2002), \texttt{hep-ex/0111091}

\bibitem{Adler:2002hy}
S.S. Adler et~al. (E787 Collaboration), Phys.Lett. \textbf{B537}, 211 (2002),
  \texttt{hep-ex/0201037}

\bibitem{Anisimovsky:2004hr}
V.~Anisimovsky et~al. (E949 Collaboration), Phys.Rev.Lett. \textbf{93}, 031801
  (2004), \texttt{hep-ex/0403036}

\bibitem{fortheNA62:2013jsa}
M.~Moulson (NA62 Collaboration) (2013), \texttt{1310.7816}

\bibitem{Christ:2015pwa}
N.H. Christ, X.~Feng, G.~Martinelli, C.T. Sachrajda, Phys. Rev. \textbf{D91},
  114510 (2015), \texttt{1504.01170}

\bibitem{Cirigliano:2011ny}
V.~Cirigliano, G.~Ecker, H.~Neufeld, A.~Pich, J.~Portoles, Rev. Mod. Phys.
  \textbf{84}, 399 (2012), \texttt{1107.6001}

\bibitem{DAmbrosio:1998gur}
G.~D'Ambrosio, G.~Ecker, G.~Isidori, J.~Portoles, JHEP \textbf{08}, 004 (1998),
  \texttt{hep-ph/9808289}

\bibitem{Batley:2009aa}
J.R. Batley et~al. (NA48/2), Phys. Lett. \textbf{B677}, 246 (2009),
  \texttt{0903.3130}

\bibitem{Batley:2011zz}
J.R. Batley et~al. (NA48/2), Phys. Lett. \textbf{B697}, 107 (2011),
  \texttt{1011.4817}

\bibitem{Tomii:2017lyo}
M.~Tomii (2017), \texttt{1706.03099}

\bibitem{Lawson}
A.~Lawson et~al., \emph{{Update on $\varepsilon_K$ with lattice QCD inputs}},
  in \emph{Proceedings, \href{http://inspirehep.net/record/1425631}{35th
  International Symposium on Lattice Field Theory (Lattice2017)}: Granada,
  Spain}, to appear in EPJ Web Conf.

\bibitem{Buchalla:2003sj}
G.~Buchalla, G.~D'Ambrosio, G.~Isidori, Nucl. Phys. \textbf{B672}, 387 (2003),
  \texttt{hep-ph/0308008}

\bibitem{Isidori:2004rb}
G.~Isidori, C.~Smith, R.~Unterdorfer, Eur. Phys. J. \textbf{C36}, 57 (2004),
  \texttt{hep-ph/0404127}

\bibitem{Lehner:2015jga}
C.~Lehner, E.~Lunghi, A.~Soni, Phys. Lett. \textbf{B759}, 82 (2016),
  \texttt{1508.01801}

\end{thebibliography}

\end{document}